\documentclass[lettersize,journal]{IEEEtran}
\usepackage{amsmath,amsfonts}
\usepackage{booktabs,tabularx}
\usepackage{algorithmic}
\usepackage{algorithm}
\usepackage{array}
\usepackage[caption=false,font=normalsize,labelfont=sf,textfont=sf]{subfig}
\usepackage{textcomp}
\usepackage{stfloats}
\usepackage{url}
\usepackage{verbatim}
\usepackage{graphicx}
\usepackage{cite}
\usepackage[colorlinks=true, linkcolor=blue, citecolor=blue, urlcolor=black]{hyperref}
\usepackage{etoolbox}

\begin{document}

\title{Learning-Assisted Beamforming Optimization for Hybrid XL-RIS-Aided Communication}

\author{Awuni Mohammed Abdul-Razak and Kyungchun Lee,~\IEEEmembership{Senior Member, IEEE}
\thanks{This work was supported in part by the Basic Science Research Program through the National Research Foundation of Korea (NRF) funded by the Ministry of Education under Grant RS-2019-NR040071 and in part by the NRF grant funded by the Korean Government (MSIT)(RS-2025-21812978). (Corresponding author: Kyungchun Lee.)}
\thanks{The authors are with the Department of Electrical and Information Engineering and the Research Center for Electrical and Information Technology, Seoul National University of Science and Technology, Seoul 01811, Republic of Korea (e-mail: armawuni24@seoultech.ac.kr, kclee@seoultech.ac.kr).}
}


\maketitle

\begin{abstract}
Hybrid active--passive extremely large-scale reconfigurable intelligent surface (XL-RIS) architectures mitigate the double-hop path loss of passive RIS while enabling near-field spatial energy focusing, but impose a computationally intensive mixed discrete--continuous beamforming design problem under strict radio-frequency (RF) power constraints. This paper investigates a hybrid XL-RIS-assisted downlink multi-user multiple-input single-output (MU-MISO) system in the near-field regime. The joint optimization of the base station (BS) precoder, discrete XL-RIS phase shifts, and active-element amplification gains is addressed via a discrete coordinate ascent (DCA)-based alternating optimization (AO) framework, in which discrete variables are updated through local index-domain searches and the BS precoder is derived in closed form via a noise-aware minimum mean-square error (MMSE) criterion. To reduce the per-sample computational cost for XL-RIS apertures, a deep unfolding (DU) framework is proposed as a low-complexity alternative, unrolling a fixed number of Riemannian gradient ascent steps on the sum-rate objective and recovering the discrete phase configuration via quantized phasor projection, with one learnable step size per layer trained in a self-supervised manner. Numerical results under near-field spherical-wave channel models confirm that the DU framework achieves sum-rate performance close to the AO scheme with substantially lower and nearly constant runtime, and that both schemes consistently outperform manifold optimization under discrete hardware constraints.
\end{abstract}

\begin{IEEEkeywords}
Near-field communications, extremely large-scale RIS (XL-RIS), hybrid active--passive RIS, MU-MISO systems, discrete beamforming optimization, deep unfolding.
\end{IEEEkeywords}

\section{Introduction}
\IEEEPARstart{T}{he} rapid growth of data-intensive wireless services—including immersive multimedia, large-scale Internet-of-Things (IoT) deployments, and intelligent autonomous systems—has placed increasing demands on modern communication networks. Reconfigurable intelligent surfaces (RIS) have emerged as a promising technology for enhancing wireless coverage and spectral efficiency by enabling programmable control of electromagnetic wave propagation. By adjusting the reflection coefficients of many low-cost passive elements, a RIS can reshape the wireless environment without additional active radio-frequency (RF) chains, thereby providing an energy-efficient alternative to conventional infrastructure-based
solutions~\cite{wuSmartReconfigurableEnvironment2020,huangReconfigurableIntelligentSurfaces2019}.
Most existing RIS studies consider far-field plane-wave propagation and purely passive phase-shift control~\cite{basarWirelessCommunicationsReconfigurable2019}.
As the physical aperture of a surface scales to the extremely large-scale RIS (XL-RIS) regime, the system enters the near-field Fresnel region, where element-specific distance variations and spherical-wave propagation must be explicitly accounted for~\cite{shenMultiBeamDesignNearField2023,gongNearFieldChannelModeling2024a,Zhang2022Holo,zhangMultiResolutionCodebookDesign2026}.
Although near-field operation enables highly focused spatial beamforming, passive XL-RIS systems remain susceptible to the severe double-hop attenuation inherent in cascaded base station (BS)--RIS--user links.
To mitigate this attenuation, hybrid active--passive RIS architectures have been proposed, in which a subset of elements integrates low-power amplification circuitry while the remaining elements operate
passively~\cite{juBeamformingOptimizationHybrid2024,nguyenHybridRelayReflectingIntelligent2022,hajjajHybridIntelligentReflecting2024}.
The active elements introduce controllable signal amplification that enhances the reflected signal power, but the amplification gain must be carefully constrained by hardware limits and a strict RF output power budget at the RIS.
The joint optimization of the BS precoder and the XL-RIS configuration under near-field conditions and discrete hardware alphabets constitutes a computationally intensive mixed-integer non-convex problem, motivating the use of learning-based methods. Various learning-based approaches have been explored to efficiently infer near-optimal RIS configurations, including deep reinforcement learning and supervised deep neural
networks~\cite{liDeepReinforcementLearningBased2023,xuRobustDeepLearningBased2022,chenDeepLearningBasedJoint2024}.

\subsection{Related Works}
\label{subsec:related_works}

Initial studies on RIS demonstrated that surface phase tuning---while maintaining near-unit reflection amplitudes---can effectively direct incident signals, thereby improving coverage, mitigating interference, and enhancing system energy efficiency. The concept of a smart radio environment
in~\cite{wuIntelligentReflectingSurfaceAided2021} demonstrated that
software-defined reflections can significantly enhance network performance.
Similarly,~\cite{wuIntelligentSurfacesEmpowered2024,hassounaSurveyReconfigurableIntelligent2023} established theoretical frameworks for passive RIS-assisted systems, demonstrating superior energy efficiency over traditional active-relaying configurations. However, these studies rely on idealized assumptions such as continuous phase control and far-field plane-wave propagation, which become increasingly restrictive for large-aperture RIS deployments, particularly when accounting for practical hardware constraints and near-field propagation.

The far-field (Fraunhofer) approximation is increasingly inadequate for characterizing signal propagation when the RIS physical dimensions are large relative to the operating wavelength. In such scenarios, the system operates within the Fresnel region, where spherical-wavefront propagation and element-wise distance variations must be explicitly modeled. A significant body of research has examined the near-field regime in the context of XL-RIS
and holographic multiple-input multiple-output (MIMO) systems. For
instance,~\cite{zengRecentAdvancesNearField2025},~\cite{chenCrossFarNearField2026},~\cite{yangNearFieldChannelEstimation2024} demonstrate that element-wise channel modeling facilitates precise beam focusing in the near-field, while~\cite{gongHolographicMIMOCommunications2024} investigates holographic MIMO architectures based on continuous-aperture representations. Furthermore,~\cite{cuiNearFieldMIMOCommunications2023},~\cite{yuanNearFieldTrackingExtremely2024} underscore the importance of spherical-wave modeling for accurate performance evaluation in electrically large arrays. These studies collectively show that near-field operation enables highly localized spatial energy concentration with finer resolution than conventional plane-wave models. However, the transition to the near-field regime introduces significant technical challenges, including the breakdown of simplified array factorizations and increased interdependence between elements, which substantially expands the design space and heightens computational complexity in XL-RIS configurations.

While passive RIS designs are inherently energy-efficient, they are
susceptible to the severe double-hop path loss in cascaded BS--RIS--user links, which degrades the received signal power and becomes a critical bottleneck in large-scale deployments or obstructed environments. To mitigate this, hybrid active--passive RIS architectures have been proposed, in which a subset of elements is equipped with power-amplification circuitry while the remainder operate passively~\cite{juBeamformingOptimizationHybrid2024,nguyenHybridRelayReflectingIntelligent2022,hajjajHybridIntelligentReflecting2024}.
Selective amplification can significantly enhance signal strength and improve interference suppression, as demonstrated
in~\cite{luLowOverheadSeparateChannel2024},~\cite{huangHybridActivePassiveRIS2024}.
Furthermore, the weighted sum-rate optimization
in~\cite{guoWeightedSumRateMaximization2020} shows that active components yield substantial performance gains in interference-limited regimes, while
\cite{alnagarEnergyEfficientHybridActive2025} investigates hybrid
relay-reflecting designs that unify amplification and reflection. Despite these advantages, the inclusion of active elements introduces significant modeling and optimization challenges: the amplifier noise must be rigorously characterized and the total RF output power must adhere to stringent hardware and power budget constraints. To date, most research on hybrid RIS has focused on far-field propagation, continuous control variables, or moderate surface dimensions, and thus these frameworks are not directly applicable to hybrid XL-RIS systems in the near-field regime under discrete phase and gain constraints.

The joint optimization of BS precoding and XL-RIS reflection coefficients constitutes a highly challenging non-convex problem due to the bilinear coupling between transmit and reflection variables. Several classical techniques have been employed, including alternating optimization~(AO), semidefinite relaxation~(SDR), fractional programming~(FP), successive convex approximation~(SCA), and manifold optimization~(MO)~\cite{alnagarEnergyEfficientHybridActive2025,maLowComplexityAlgorithmicFramework2020}. Among these, AO-based frameworks have gained considerable attention due to their flexibility and robust empirical performance. However, discrete phase shifts and quantized amplification gains transform the problem into a combinatorial search whose dimensionality renders traditional approaches computationally prohibitive as the number of XL-RIS elements grows. While MO-based methods exhibit favorable convergence on the unit-modulus manifold, their effectiveness diminishes under stringent discrete hardware constraints and RF power budgets. The addition of gain-selection variables for active elements further introduces mixed-integer components that significantly increase the problem size.

Recent learning-based methods have demonstrated considerable potential in mitigating these scalability limitations. For instance,
\cite{sheenDeepLearningBased2021} validates deep learning for RIS phase design, while~\cite{fengDeepReinforcementLearning2025,nguyenDeepReinforcementLearning2021a} explore reinforcement learning for adaptive RIS configuration. Tailored neural network architectures for phase prediction are investigated
in~\cite{fondo-ferreiroNeuralNetworksPhase2025}, and multi-agent
reinforcement learning (MARL) for distributed RIS control is examined
in~\cite{zhuMultiAgentReinforcementLearningBased2024}.

Despite these advances, most existing studies address passive RIS
architectures under far-field assumptions and adopt global learning strategies that infer the complete RIS configuration from high-dimensional channel state information. In near-field XL-RIS systems, strong interdependence among elements and the exponential growth of the discrete action space significantly complicate such global paradigms. As noted in~\cite{liPhaseShiftDesign2022}, there remains a pronounced gap in scalable, learning-assisted frameworks that explicitly account for hybrid XL-RIS architectures, discrete phase and gain constraints, amplifier-induced noise, and near-field propagation.

\subsection{Motivations and Contributions}

Motivated by the scalability challenges of optimizing hybrid XL-RIS systems in the near-field regime, this work develops a low-complexity beamforming framework for downlink MU-MISO transmission. Although the proposed framework is not restricted to XL-RIS deployments and can be applied to more general RIS architectures, it is specifically designed to address the high-dimensional optimization challenges that become particularly significant in XL-RIS systems.
A physically consistent spherical-wave channel model is adopted, and the joint beamforming design is formulated as a sum-rate maximization problem subject to discrete phase control, quantized amplification gains, and strict RF power constraints on the active XL-RIS elements. A DCA-based AO framework is first developed to solve the resulting mixed discrete--continuous optimization problem under practical hardware constraints.
To reduce the per-sample computational cost of iterative
optimization in large-scale XL-RIS deployments, a low-complexity deep unfolding (DU) framework is subsequently developed as an optimization-driven alternative, unfolding a fixed number of gradient-based update steps directly on the physical sum-rate objective.

The main contributions of this paper are summarized as follows:

\begin{itemize}

    \item We develop a comprehensive system model for hybrid active--passive XL-RIS-assisted MU-MISO transmission in the near-field regime. The model captures spherical-wave propagation, discrete phase shifts, quantized amplification gains, amplifier-induced noise, and practical RF power constraints on the active XL-RIS elements. Based on this model, we formulate a sum-rate maximization problem that jointly optimizes the BS precoder, the discrete XL-RIS phase shifts, and the amplification gains. To address the resulting mixed discrete--continuous non-convex problem, we propose a coordinate-ascent AO framework in which discrete updates are performed via local index-domain searches over the phase and gain codebooks, thereby avoiding exhaustive enumeration. Multiple random restarts are employed to improve robustness against local optima, and the BS precoder is updated in closed form via a noise-aware MMSE solution.

    \item We propose a lightweight DU framework as a low-complexity
    alternative to the DCA-based AO algorithm for hybrid XL-RIS phase optimization. The framework unrolls a fixed number of Riemannian gradient ascent steps on the sum-rate objective over the continuous unit-phasor domain, retracting onto the unit-modulus manifold at each layer. A single positive step size per layer is learned, yielding a parameter count equal to the number of layers independent of system dimensions. The discrete phase configuration is recovered via quantized phasor projection, with a straight-through estimator (STE) enabling gradient propagation through the non-differentiable quantization step. The framework is trained self-supervisedly by directly maximizing the  quantized sum-rate, requiring no labeled data.
    
    \item Extensive simulations are conducted to evaluate the
    proposed framework. The results demonstrate that the DU framework achieves sum-rate performance close to that of the AO scheme while significantly reducing computational runtime, and consistently outperforms the MO baseline under discrete hardware constraints.

\end{itemize}

The remainder of this paper is organized as follows.
Section~\ref{sec:system_model} presents the system and channel models. Section~\ref{sec:problem_formulation} formulates the joint optimization problem. Section~\ref{sec:proposed_method} describes the proposed AO algorithm and DU framework. Section~\ref{sec:complexity_analysis} analyzes
the computational complexity of the considered schemes.
Section~\ref{sec:simulation_setup} presents the simulation results, and Section~\ref{sec:conclusion} concludes the paper.

\textit{Notations:}
Matrices and vectors are denoted by boldface upper-case and boldface lower-case letters, respectively. The operators $(\cdot)^{\mathrm{T}}$ and $(\cdot)^{\mathrm{H}}$ denote the transpose and conjugate transpose, respectively. The operators $\Re\{\cdot\}$ and $\Im\{\cdot\}$ denote the real and imaginary parts of a complex quantity, respectively.
The operator $\mathrm{diag}(a_1,\ldots,a_N)$ denotes a diagonal matrix with entries $a_1,\ldots,a_N$. The magnitude of a scalar is denoted by $|\cdot|$, while $\|\cdot\|_2$ and $\|\cdot\|_F$ denote the Euclidean norm of a vector and the Frobenius norm of a matrix, respectively. The sets of real and complex numbers are denoted by $\mathbb{R}$ and $\mathbb{C}$, respectively.
For a vector $\mathbf{a}$, $[\mathbf{a}]_n$ denotes its $n$-th element; for a matrix $\mathbf{A}$, $[\mathbf{A}]_{m,n}$ denotes the $(m,n)$-th entry, while $[\mathbf{A}]_{k,:}$ and $[\mathbf{A}]_{:,k}$ denote its $k$-th row and $k$-th column, respectively. The expectation operator is denoted by $\mathbb{E}\{\cdot\}$, and $\mathcal{CN}(\mu,\sigma^2)$ denotes the circularly symmetric complex Gaussian distribution with mean $\mu$ and variance $\sigma^2$. The unit circle is denoted by $\mathbb{S}^1$, and $(\mathbb{S}^1)^{N_r}$ denotes its $N_r$-fold Cartesian product, which constitutes the continuous phasor manifold of the XL-RIS.

\section{System Model}
\label{sec:system_model}

We consider a downlink MU-MISO communication system assisted by a hybrid XL-RIS operating in the near-field regime. The BS, equipped with $M$ transmit antennas, simultaneously serves $K$ single-antenna user equipments (UEs). Communication between the BS and the UEs is realized exclusively through the XL-RIS, while the direct BS--UE link is assumed to be either blocked or severely attenuated and is therefore considered negligible.

\subsection{System Architecture}

\begin{figure}[!t]
\centering
    \includegraphics[width=0.95\columnwidth]{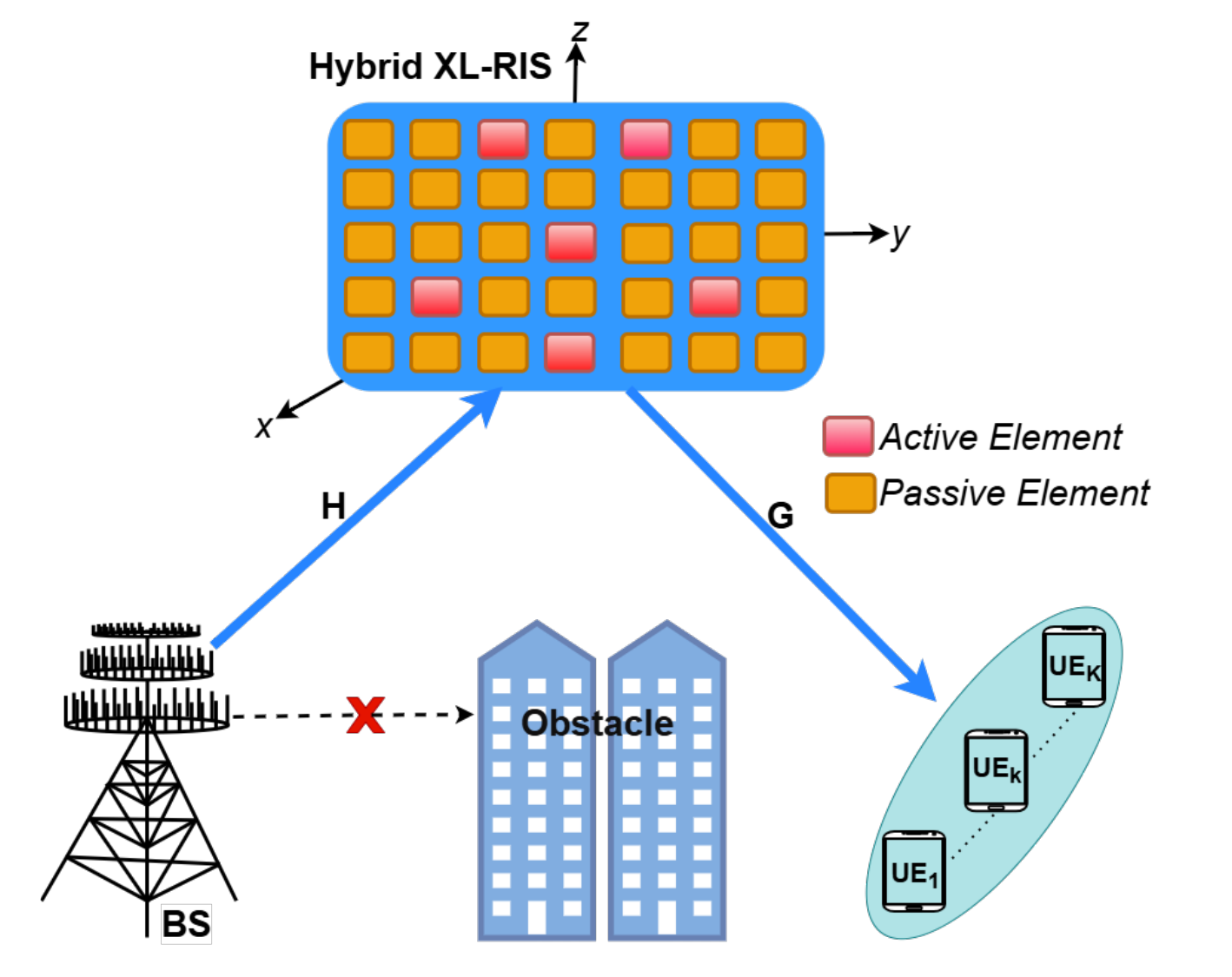}
    \caption{Hybrid XL-RIS-assisted downlink MU-MISO system architecture operating in the near-field regime.}
    \label{fig:system_model}
\end{figure}

The hybrid XL-RIS-assisted downlink MU-MISO architecture is illustrated in Fig.~\ref{fig:system_model}. The XL-RIS comprises a large number of reflecting elements, a subset of which is equipped with amplification capability, while the remaining elements operate in passive reflection mode.
All XL-RIS elements implement discrete phase shifts, and the active elements additionally apply quantized amplification gains.

The system is subject to two power constraints: a transmit power constraint at the BS and a strict RF output power budget at the XL-RIS. Owing to the electrically large aperture of the XL-RIS and the moderate propagation distances, the signal propagates in the Fresnel (near-field) region, where spherical-wave propagation characteristics dominate and must be explicitly incorporated into the channel model.

\subsection{Hybrid XL-RIS Structure and Transmit Signal}

The XL-RIS comprises $N_r$ reflecting elements, including $N_a$ active elements and $N_p$ passive elements, with $N_r = N_a + N_p$. The complete index set is denoted by $\mathcal{N} = \{1,\ldots,N_r\}$ and partitioned into two disjoint subsets: the active set $\mathcal{A}$ and the passive set
$\mathcal{P}$, satisfying $\mathcal{A} \cap \mathcal{P} = \emptyset$.
The BS adopts linear precoding with
$\mathbf{W} = [\mathbf{w}_1,\ldots,\mathbf{w}_K] \in \mathbb{C}^{M \times K}$ to transmit independent unit-power symbols $\{s_k\}$, where $\mathbb{E}\{|s_k|^2\} = 1$. The transmitted signal is
\begin{equation}
    \mathbf{x} = \sum_{k=1}^{K} \mathbf{w}_k s_k,
\end{equation}
subject to the transmit power constraint
\begin{equation}
    \label{eq:tx_power_constraint}
    \|\mathbf{W}\|_F^2 \le P_{\mathrm{TX}}.
\end{equation}

\subsection{Near-Field Channel Model}

To account for the Fresnel region, a spherical-wave model is adopted for both the BS--XL-RIS and XL-RIS--UE links. Specifically, the BS--XL-RIS channel is represented by $\mathbf{H} \in \mathbb{C}^{N_r \times M}$, whose $(n,m)$-th entry is given by~\cite{article,luTutorialNearFieldXLMIMO2024}
\begin{equation}
    \label{eq:H_channel}
    H_{n,m}
    =
    \sum_{p=1}^{P} \sum_{\ell=1}^{L_p}
    \alpha^{\mathrm{BR}}_{p,\ell}
    \frac{\lambda}{4\pi d^{\mathrm{BR}}_{p,\ell,n,m}}
    \exp\!\left(-j\frac{2\pi}{\lambda} d^{\mathrm{BR}}_{p,\ell,n,m}\right),
\end{equation}
where $d^{\mathrm{BR}}_{p,\ell,n,m}$ denotes the propagation distance between the $m$-th BS antenna and the $n$-th XL-RIS element for the $(p,\ell)$-th subpath, $\alpha^{\mathrm{BR}}_{p,\ell}$ is the corresponding complex path gain, $\lambda$ is the carrier wavelength, $P$ is the number of propagation clusters, and $L_p$ is the number of subpaths within the $p$-th cluster.

Similarly, the XL-RIS--UE channel is described by
$\mathbf{G} \in \mathbb{C}^{K \times N_r}$, whose $(k,n)$-th entry is
\begin{equation}
    \label{eq:G_channel}
    g_{k,n}
    =
    \sum_{p=1}^{P} \sum_{\ell=1}^{L_p}
    \beta^{\mathrm{RU}}_{p,\ell}
    \frac{\lambda}{4\pi d^{\mathrm{RU}}_{p,\ell,n,k}}
    \exp\!\left(-j\frac{2\pi}{\lambda} d^{\mathrm{RU}}_{p,\ell,n,k}\right),
\end{equation}
where $d^{\mathrm{RU}}_{p,\ell,n,k}$ denotes the propagation distance between the $n$-th XL-RIS element and the $k$-th user for the $(p,\ell)$-th subpath, and $\beta^{\mathrm{RU}}_{p,\ell}$ is the corresponding complex path gain. The $k$-th row of $\mathbf{G}$ is denoted by $\mathbf{g}_k^{\mathrm{T}} \in \mathbb{C}^{1 \times N_r}$, where $\mathbf{g}_k \in \mathbb{C}^{N_r}$ is the channel vector between the XL-RIS and user~$k$.

\subsection{Hybrid XL-RIS Configuration}

Each XL-RIS element applies a complex reflection coefficient
$\psi_n = \gamma_n e^{j\theta_n}$, where $\theta_n$ denotes the phase shift
and $\gamma_n$ the reflection amplitude (or amplification gain) of the $n$-th
element.
The phase shift $\theta_n$ is selected from the discrete set
\begin{equation}
    \label{eq:phase_set}
    \mathcal{F} =
    \left\{
    0,\, \frac{2\pi}{L_{\theta}},\, \ldots,\,
    \frac{2\pi(L_{\theta}-1)}{L_{\theta}}
    \right\},
    \qquad L_{\theta} = 2^B,
\end{equation}
where $B$ denotes the phase resolution in bits. For passive XL-RIS elements, the reflection amplitude satisfies $\gamma_n = 1$. For active elements, the amplification gain $\gamma_n$ is selected from a discrete gain set $\mathcal{G} \subset [1,\gamma_{\max}]$.
Let $\boldsymbol{\Theta} \in \mathbb{C}^{N_r \times N_r}$ denote the diagonal phase-shift matrix
\begin{equation}
    \boldsymbol{\Theta}
    =
    \mathrm{diag}\!\left(
    e^{j\theta_1},\, e^{j\theta_2},\, \ldots,\, e^{j\theta_{N_r}}
    \right),
\end{equation}
and let $\boldsymbol{\Gamma} \in \mathbb{R}^{N_r \times N_r}$ denote the
diagonal gain matrix
\begin{equation}
    \boldsymbol{\Gamma}
    =
    \mathrm{diag}\!\left(
    \gamma_1,\, \gamma_2,\, \ldots,\, \gamma_{N_r}
    \right).
\end{equation}
The overall reflection matrix of the hybrid XL-RIS is therefore
\begin{equation}
    \label{eq:Psi}
    \boldsymbol{\Psi}
    =
    \boldsymbol{\Gamma}\boldsymbol{\Theta}
    =
    \mathrm{diag}\!\left(
    \gamma_1 e^{j\theta_1},\,
    \gamma_2 e^{j\theta_2},\,
    \ldots,\,
    \gamma_{N_r} e^{j\theta_{N_r}}
    \right)
    \in \mathbb{C}^{N_r \times N_r}.
\end{equation}

\subsection{Effective Channel and Noise Modeling}

The cascaded BS--UE channel under hybrid XL-RIS operation is
\begin{equation}
    \label{eq:Heff}
    \mathbf{H}_{\mathrm{eff}}
    =
    \mathbf{G}\boldsymbol{\Psi}\mathbf{H}
    \in \mathbb{C}^{K \times M}.
\end{equation}
Thermal noise at user $k$ is modeled as $n_k \sim \mathcal{CN}(0,\sigma_k^2)$.
Active XL-RIS elements introduce amplification noise; let
$\mathbf{z}_a \in \mathbb{C}^{N_r}$ denote the amplification noise vector
with entries
\begin{equation}
    \label{eq:amp_noise}
    [\mathbf{z}_a]_n =
    \begin{cases}
    \mathcal{CN}(0,\sigma_a^2), & n \in \mathcal{A},\\
    0,                           & n \notin \mathcal{A},
    \end{cases}
\end{equation}
where $\sigma_a^2$ is the amplifier noise power at each active element.
After reflection, the amplification noise propagates through the XL-RIS--UE link, yielding the effective noise variance at user $k$:
\begin{equation}
    \label{eq:sigma_tilde}
    \tilde{\sigma}_k^2
    =
    \sigma_k^2
    +
    \sigma_a^2 \sum_{n \in \mathcal{A}} \gamma_n^2\, |g_{k,n}|^2.
\end{equation}
The effective noise covariance matrix is
$\mathbf{R} = \mathrm{diag}(\tilde{\sigma}_1^2,\ldots,\tilde{\sigma}_K^2)$,
and the noise-aware MMSE precoder is
\begin{equation}
    \label{eq:W0}
    \mathbf{W}_0
    =
    \mathbf{H}_{\mathrm{eff}}^{\mathrm{H}}
    \!\left(
    \mathbf{H}_{\mathrm{eff}}\mathbf{H}_{\mathrm{eff}}^{\mathrm{H}}
    +
    \mathbf{R}
    \right)^{-1}.
\end{equation}
The precoder in~\eqref{eq:W0} corresponds to the weighted MMSE solution under non-uniform user noise covariance $\mathbf{R}$, thereby accounting for amplification-induced noise heterogeneity across users. Finally, $\mathbf{W}_0$ is scaled to satisfy the transmit power constraint in~\eqref{eq:tx_power_constraint}.

\subsection{Received Signal Model}

The received baseband signal at user $k$ is
\begin{equation}
    \label{eq:yk}
    y_k
    =
    \mathbf{g}_k^{\mathrm{T}} \boldsymbol{\Psi}\mathbf{H}\mathbf{w}_k s_k
    +
    \sum_{i \neq k} \mathbf{g}_k^{\mathrm{T}} \boldsymbol{\Psi}\mathbf{H}\mathbf{w}_i s_i
    +
    \mathbf{g}_k^{\mathrm{T}} \boldsymbol{\Psi}_{\mathcal{A}} \mathbf{z}_a
    +
    n_k,
\end{equation}
where $\boldsymbol{\Psi}_{\mathcal{A}}$ denotes the diagonal matrix retaining only the active-element entries of $\boldsymbol{\Psi}$, with passive-element entries set to zero. The SINR at user $k$ is
\begin{equation}
    \label{eq:SINR}
    \mathrm{SINR}_k
    =
    \frac{
    |\mathbf{g}_k^{\mathrm{T}} \boldsymbol{\Psi}\mathbf{H}\mathbf{w}_k|^2
    }{
    \displaystyle
    \sum_{i \neq k}
    |\mathbf{g}_k^{\mathrm{T}} \boldsymbol{\Psi}\mathbf{H}\mathbf{w}_i|^2
    +
    \sigma_a^2 \|\mathbf{g}_k^{\mathrm{T}} \boldsymbol{\Psi}_{\mathcal{A}}\|_2^2
    +
    \sigma_k^2
    },
\end{equation}
and the achievable rate of user $k$ is
$R_k = \log_2(1 + \mathrm{SINR}_k)$.

\section{Problem Formulation}
\label{sec:problem_formulation}

Based on the system model in Section~\ref{sec:system_model}, we formulate the joint beamforming design problem for the hybrid XL-RIS-assisted downlink MU-MISO system. The objective is to jointly optimize the BS precoding matrix, the XL-RIS phase shifts, and the active-element amplification gains to maximize the system sum-rate, subject to practical hardware constraints including discrete phase control, quantized amplification gains, the BS transmit power limit, and the XL-RIS RF output power budget.

The optimization involves three coupled sets of variables:
\begin{itemize}
    \item BS precoding matrix $\mathbf{W} \in \mathbb{C}^{M \times K}$,
    \item XL-RIS phase vector
          $\boldsymbol{\theta} = [\theta_1,\ldots,\theta_{N_r}]^{\mathrm{T}}$,
    \item amplification gain vector
          $\boldsymbol{\gamma} = [\gamma_1,\ldots,\gamma_{N_r}]^{\mathrm{T}}$.
\end{itemize}
The achievable downlink sum-rate is
\begin{equation}
\label{eq:sum_rate}
    R_{\mathrm{sum}}(\mathbf{W}, \boldsymbol{\theta}, \boldsymbol{\gamma})
    =
    \sum_{k=1}^{K}
    \log_2\!\left(1 + \mathrm{SINR}_k \right),
\end{equation}
where $\mathrm{SINR}_k$ is defined in~\eqref{eq:SINR}. The joint
optimization problem is formulated as
\begin{equation}
\label{prob:P0}
    \begin{aligned}
    (\mathrm{P0}) \quad
    \max_{\mathbf{W},\,\boldsymbol{\theta},\,\boldsymbol{\gamma}} \;\;
    & R_{\mathrm{sum}}(\mathbf{W}, \boldsymbol{\theta}, \boldsymbol{\gamma}) \\[1mm]
    \text{s.t.}\quad
    & \mathrm{C1{:}}\; \theta_n \in \mathcal{F},\; \forall n \in \mathcal{N}, \\[0.5mm]
    & \phantom{\mathrm{C1{:}}\;} \gamma_n = 1,\; \forall n \in \mathcal{P}, \\[0.5mm]
    & \phantom{\mathrm{C1{:}}\;} \gamma_n \in \mathcal{G},\; \forall n \in \mathcal{A}, \\[0.8mm]
    & \mathrm{C2{:}}\; \|\mathbf{W}\|_F^2 \le P_{\mathrm{TX}}, \\[0.8mm]
    & \mathrm{C3{:}}\;
    \sum_{n \in \mathcal{A}}
    \gamma_n^2 \!\left(P_{\mathrm{in},n} + \sigma_a^2\right)
    \le P_{\mathrm{RIS}}^{\max},
    \end{aligned}
\end{equation}
where
\begin{equation}
\label{eq:Pin}
    P_{\mathrm{in},n}
    =
    \|\mathbf{h}_n^{\mathrm{T}} \mathbf{W}\|_2^2
    =
    \left[\mathbf{H}\mathbf{W}\mathbf{W}^{\mathrm{H}}\mathbf{H}^{\mathrm{H}}\right]_{n,n},
    \quad \forall n \in \mathcal{N},
\end{equation}
is the incident signal power at XL-RIS element $n$, and
$\mathbf{h}_n^{\mathrm{T}} \in \mathbb{C}^{1 \times M}$ is the $n$-th row of $\mathbf{H}$.

Problem~(P0) is a mixed discrete--continuous non-convex optimization problem. Non-convexity arises from the bilinear coupling among the BS precoder, the XL-RIS phase shifts, and the amplification gains in the SINR expressions.The discrete phase and gain constraints in C1 additionally introduce combinatorial complexity, and the RF power budget in C3 couples the active-element gains across the XL-RIS.

Due to the mixed discrete--continuous and non-convex structure of~(P0), obtaining the globally optimal solution is computationally intractable for large-scale XL-RIS systems. To address this, a DCA-based AO framework is first developed to compute high-quality
locally stationary solutions under the practical hardware constraints. A low-complexity DU framework is then proposed to enable scalable XL-RIS phase optimization with substantially reduced computational complexity.

\section{Proposed Solution}
\label{sec:proposed_method}

The combinatorial growth of the discrete configuration space of~(P0) renders exhaustive search intractable for large-scale XL-RIS deployments. To address this, we develop a two-stage solution framework: a DCA-based AO algorithm that computes high-quality locally stationary solutions under the discrete hardware and RF power constraints, and a low-complexity DU framework as a computationally efficient alternative for scalable XL-RIS phase optimization.

\subsection{Alternating Optimization}
\label{subsec:ao}

The proposed AO framework alternates between (i) closed-form MMSE-based BS precoder updates and (ii) discrete coordinate-wise refinement of the XL-RIS phase shifts and active gains. All discrete hardware alphabets and the XL-RIS RF power budget are strictly enforced at every iteration through sum-rate evaluation and feasibility checks.
Specifically, let the discrete phase and gain index sets be
$\mathcal{Q} = \{0,1,\ldots,L_{\theta}-1\}$ and
$\mathcal{B} = \{0,1,\ldots,L_g-1\}$, where $L_g = |\mathcal{G}|$.
At iteration $t$, the XL-RIS configuration is represented by the
phase-index vector
$\mathbf{q}^{(t)} = [q_1^{(t)},\ldots,q_{N_r}^{(t)}]^{\mathrm{T}}$
and the active gain-index vector
$\mathbf{b}^{(t)} = [b_1^{(t)},\ldots,b_{N_a}^{(t)}]^{\mathrm{T}}$.
The corresponding phases and gains are recovered via the mappings
$\theta_n^{(t)} = 2\pi q_n^{(t)}/L_{\theta}$ and
$\gamma_n^{(t)} = \mathcal{G}[b_n^{(t)}]$, respectively, and the XL-RIS reflection matrix at iteration $t$ is
$\boldsymbol{\Psi}^{(t)} =
\mathrm{diag}(\gamma_1^{(t)}e^{j\theta_1^{(t)}},\ldots,
\gamma_{N_r}^{(t)}e^{j\theta_{N_r}^{(t)}})$.

\subsubsection{BS Precoder Update}

For a fixed XL-RIS configuration $(\boldsymbol{\theta}^{(t)},
\boldsymbol{\gamma}^{(t)})$, the effective cascaded channel
$\mathbf{H}_{\mathrm{eff}}$ remains unchanged. The BS precoder is updated via a noise-aware MMSE criterion, where $\mathbf{R}$ captures both the receiver thermal noise and the amplification noise forwarded by the active XL-RIS elements. Using the Woodbury identity, the update requires inverting only a $K\times K$ matrix, yielding the closed-form solution
\begin{equation}
    \label{eq:mmse_user_domain}
    \mathbf{W}_0^{(t+1)}
    =
    \mathbf{H}_{\mathrm{eff}}^{\mathrm{H}}
    \!\left(
    \mathbf{H}_{\mathrm{eff}}\mathbf{H}_{\mathrm{eff}}^{\mathrm{H}}
    +
    \mathbf{R}
    \right)^{\!-1}\!,
    \quad
    \mathbf{W}^{(t+1)} = \alpha\,\mathbf{W}_0^{(t+1)},
\end{equation}
where $\alpha = \sqrt{P_{\mathrm{TX}}/\|\mathbf{W}_0^{(t+1)}\|_F^2}$
ensures compliance with constraint~C2.

\subsubsection{Discrete Neighborhood Definition}

To reduce computational complexity, the XL-RIS parameters are refined via local index-domain search rather than exhaustive enumeration, following principles similar
to~\cite{farsaDiscreteCoordinateDescent2020,kimaryoLowcomplexityIRSBeamforming2023}.
For a phase index $q_n^{(t)}\in\mathcal{Q}$, the search neighborhood is
\begin{equation}
    \label{eq:phase_nbhd}
    \mathcal{N}_{\mathcal{Q}}(q_n^{(t)})
    =
    \bigl\{
    (q_n^{(t)}-1)\!\!\!\!\mod L_{\theta},\;
    (q_n^{(t)}+1)\!\!\!\!\mod L_{\theta}
    \bigr\}.
\end{equation}
For a gain index $b_n^{(t)}\in\mathcal{B}$, the search neighborhood is
\begin{equation}
    \label{eq:gain_nbhd}
    \mathcal{N}_{\mathcal{B}}(b_n^{(t)})
    =
    \bigl\{
    \max(0,\,b_n^{(t)}-1),\;
    \min(L_g-1,\,b_n^{(t)}+1)
    \bigr\}.
\end{equation}
Phase candidates are evaluated cyclically, reflecting the circular structure of $\mathcal{F}$, whereas gain candidates are clipped at the boundaries of $\mathcal{B}$ since the gain set is ordered and non-cyclic.

\subsubsection{Discrete Phase Optimization}
Given $\mathbf{W}^{(t+1)}$ and fixed gains $\mathbf{b}^{(t)}$, the phase indices are updated sequentially via discrete coordinate ascent:
\begin{equation}
    \label{eq:phase_update}
    q_n^{(t+1)}
    =
    \arg\max_{q\,\in\,\mathcal{N}_{\mathcal{Q}}(q_n^{(t)})}
    R_{\mathrm{sum}}\!\left(
    \mathbf{W}^{(t+1)},\,\mathbf{q}_{-n}^{(t)},\,q,\,\mathbf{b}^{(t)}
    \right),
\end{equation}
where $\mathbf{q}_{-n}$ denotes the phase-index vector with the $n$-th entry excluded. In this process, only strictly improving updates are accepted.

\subsubsection{Discrete Active-Gain Optimization}

For each active element $n\in\{1,\ldots,N_a\}$, the gain
index is updated analogously:
\begin{equation}
    \label{eq:gain_update}
    b_n^{(t+1)}
    =
    \arg\max_{b\,\in\,\mathcal{N}_{\mathcal{B}}(b_n^{(t)})}
    R_{\mathrm{sum}}\!\left(
    \mathbf{W}^{(t+1)},\,\mathbf{q}^{(t+1)},\,\mathbf{b}_{-n}^{(t)},\,b
    \right),
\end{equation}
subject to constraint~C3 in~(P0), where $\mathbf{b}_{-n}$ denotes the gain-index vector with the $n$-th entry excluded. Only strictly improving updates are accepted. For any candidate that violates the RF power constraint, the corresponding gain is reduced via downward quantization until $P_{\mathrm{RIS}} \le P_{\mathrm{RIS}}^{\max}$, ensuring all accepted updates remain feasible.

\subsubsection{Multiple Restarts and AO Procedure}

Algorithm~\ref{alg:ao_local_search} summarizes the overall AO procedure. Since (P0) is non-convex, the discrete coordinate ascent may converge to different local optima depending on initialization. Multiple random restarts are therefore employed, with the phase and gain indices drawn randomly from $\mathcal{Q}$ and $\mathcal{B}$ at each restart, subject to C3. Within each restart, iterations proceed until no further improvement is observed or the maximum iteration count $T$ is reached. Since every coordinate update performs a local maximization, the objective is non-decreasing across iterations, and convergence in a finite number of steps is guaranteed by the finiteness of the discrete feasible set. After all $R$ restarts, the configuration achieving the highest objective value is returned.
Although each XL-RIS element is optimized individually,
its reflection coefficient combines coherently with those of all other elements in the cascaded channel, so that a single-element update affects the aggregate received signal and the system sum-rate.

\begin{algorithm}[!t]
\small
\caption{DCA-Based AO for Hybrid XL-RIS Configuration}
\label{alg:ao_local_search}
\begin{algorithmic}[1]
\REQUIRE $\mathbf{H},\,\mathbf{G}$;\;
         $\mathcal{Q},\,\mathcal{B}$;\; $\mathcal{F},\,\mathcal{G}$;\;
         $P_{\mathrm{TX}},\,P_{\mathrm{RIS}}^{\max}$;\; $T,\,R$
\ENSURE  $\mathbf{q}^\star,\,\mathbf{b}^\star,\,
         (\boldsymbol{\theta}^\star,\boldsymbol{\gamma}^\star),\,
         \mathbf{W}^\star,\,R_{\mathrm{sum}}^\star$
\STATE $R_{\mathrm{sum}}^\star \leftarrow -\infty$
\FOR{$r = 1$ \TO $R$}
    \STATE Randomly initialize $\mathbf{q},\,\mathbf{b}$ subject to C3
    \FOR{$t = 1$ \TO $T$}
        \STATE Map $(\mathbf{q},\mathbf{b})\to(\boldsymbol{\theta},
               \boldsymbol{\gamma})$;
               construct $\boldsymbol{\Psi}$;
               compute $\mathbf{H}_{\mathrm{eff}}$
        \STATE Update $\mathbf{W}$ via~\eqref{eq:mmse_user_domain}
        \STATE $R_{\mathrm{cur}} \leftarrow
               R_{\mathrm{sum}}(\mathbf{W},\mathbf{q},\mathbf{b})$
        \FOR{$n = 1$ \TO $N_r$}
            \FOR{$q \in \mathcal{N}_{\mathcal{Q}}(q_n)$}
                \STATE $\tilde{R} \leftarrow
                       R_{\mathrm{sum}}(\mathbf{W},\mathbf{q}_{-n},q,\mathbf{b})$
                \IF{$\tilde{R} > R_{\mathrm{cur}}$}
                    \STATE $q_n \leftarrow q$;\;
                           $R_{\mathrm{cur}} \leftarrow \tilde{R}$
                \ENDIF
            \ENDFOR
        \ENDFOR
        \FOR{$n = 1$ \TO $N_a$}
            \FOR{$b \in \mathcal{N}_{\mathcal{B}}(b_n)$}
                \STATE Form $\tilde{\mathbf{b}}$; project to satisfy
                       $P_{\mathrm{RIS}} \le P_{\mathrm{RIS}}^{\max}$
                \STATE $\tilde{R} \leftarrow
                       R_{\mathrm{sum}}(\mathbf{W},\mathbf{q},\tilde{\mathbf{b}})$
                \IF{$\tilde{R} > R_{\mathrm{cur}}$}
                    \STATE $b_n \leftarrow \tilde{b}_n$;\;
                           $R_{\mathrm{cur}} \leftarrow \tilde{R}$
                \ENDIF
            \ENDFOR
        \ENDFOR
        \IF{no improvement}
            \STATE \textbf{break}
        \ENDIF
    \ENDFOR
    \IF{$R_{\mathrm{cur}} > R_{\mathrm{sum}}^\star$}
        \STATE $(\mathbf{q}^\star,\mathbf{b}^\star,\mathbf{W}^\star)
               \leftarrow (\mathbf{q},\mathbf{b},\mathbf{W})$;\;
               $R_{\mathrm{sum}}^\star \leftarrow R_{\mathrm{cur}}$
    \ENDIF
\ENDFOR
\end{algorithmic}
\end{algorithm}

\subsection{Deep Unfolding for Low-Complexity XL-RIS Phase Optimization}
\label{subsec:du}

\begin{figure}
    \centering
    \includegraphics[width=0.95\columnwidth]{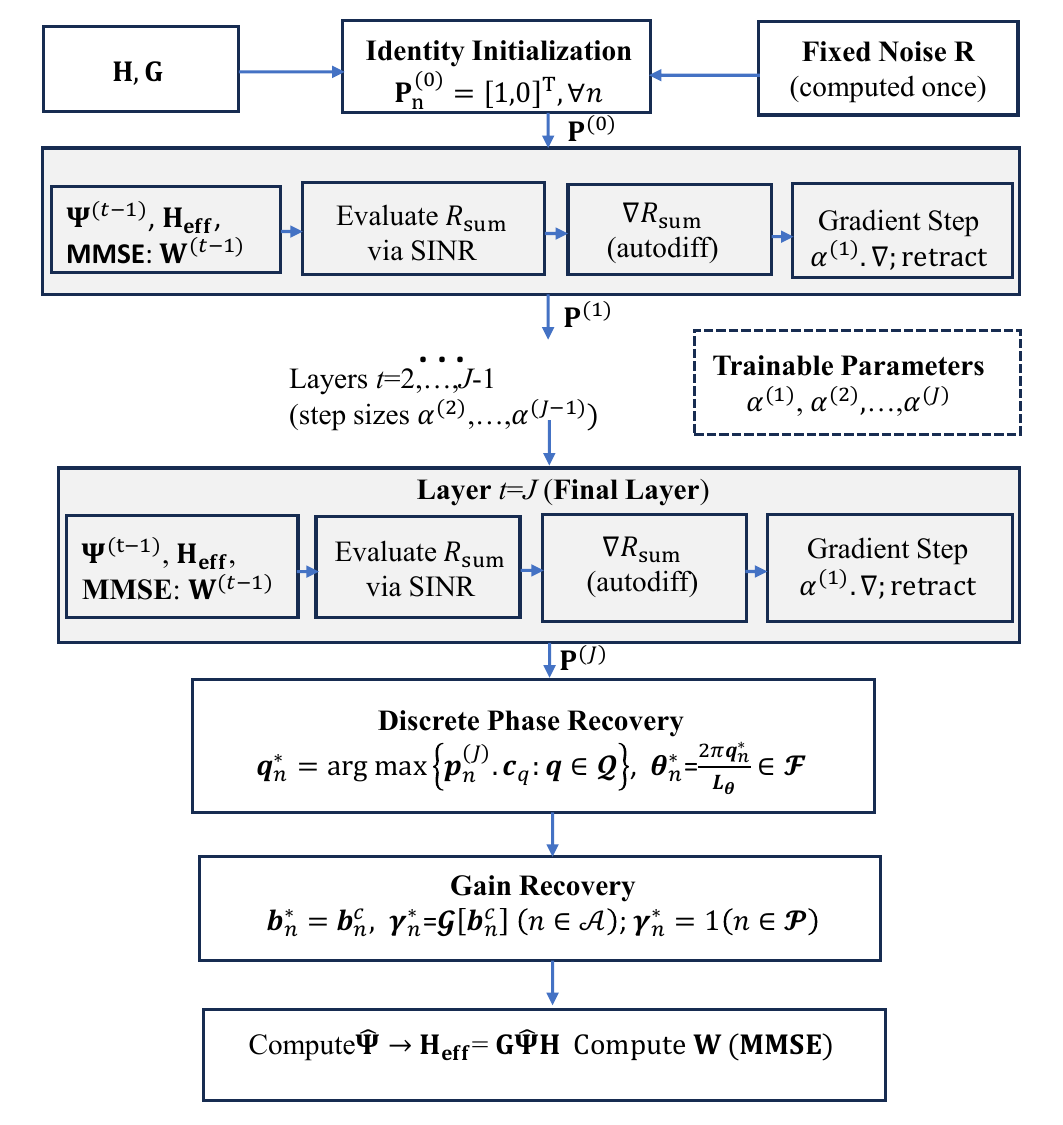}
    \caption{Architecture of the proposed DU framework.
             The phasor state $\mathbf{P}^{(0)}$ is initialized at the identity phase configuration and updated across $J$ Riemannian gradient ascent layers, each parameterized by a single learnable step size $\alpha^{(t)}$. After the final layer, the continuous phasors are projected onto the discrete codebook $\mathcal{F}$ to recover the phase-index vector $\mathbf{q}^*$, and the active-element gains are assigned the conservative discrete level $\boldsymbol{\gamma}^*$. The BS precoder $\mathbf{W}$ is then computed via the noise-aware MMSE criterion.}
    \label{fig:du_architecture}
\end{figure}

Although the AO algorithm in Section~\ref{subsec:ao} achieves near-optimal sum-rate performance, its iterative discrete coordinate updates incur substantial computational complexity for XL-RIS deployments.
To address this, we propose a DU framework that unfolds a fixed number of Riemannian gradient ascent steps on the sum-rate objective into a fixed-depth computational architecture consisting of $J$ gradient ascent layers, each with a learnable positive step size. The active-element gains are fixed throughout the unfolding, and the continuous phasors produced at the final layer are quantized to the discrete codebook $\mathcal{F}$.
Since $J \ll R\,T_{\mathrm{AO}}$, the DU framework achieves substantially lower computational complexity than the AO algorithm. See Fig.~\ref{fig:du_architecture} that describes the fixed-depth computational network.

\subsubsection{Continuous Phasor Representation}

Discrete phase indices in $\mathcal{Q}$ are non-differentiable, preventing gradient computation through the quantization step. The DU framework therefore operates on a continuous relaxation: the phase of each XL-RIS element is represented as a two-dimensional unit phasor on the unit circle $\mathbb{S}^1$. At unrolling step $t$, the phase state of the full $N_r$-element array is collected in the matrix
\begin{equation}
\label{eq:phasor_matrix}
    \mathbf{P}^{(t)} \;=\;
    \begin{bmatrix}
        \cos\theta_1^{(t)}     & \sin\theta_1^{(t)}     \\
        \vdots                  & \vdots                  \\
        \cos\theta_{N_r}^{(t)} & \sin\theta_{N_r}^{(t)}
    \end{bmatrix}
    \in \mathbb{R}^{N_r \times 2},
\end{equation}
where the $n$-th row $\mathbf{p}_{n}^{(t)} =
[\cos\theta_n^{(t)},\;\sin\theta_n^{(t)}]^{\mathrm{T}} \in \mathbb{R}^2$
is the unit phasor of element $n$ at step $t$, satisfying
$\|\mathbf{p}_{n}^{(t)}\|_2 = 1$. The corresponding complex reflection
coefficient is
\begin{equation}
\label{eq:complex_phasor}
    v_n^{(t)} \;=\; p_{n,1}^{(t)} \;+\; j\,p_{n,2}^{(t)}
              \;=\; e^{j\theta_n^{(t)}},
\end{equation}
where $p_{n,1}^{(t)}$ and $p_{n,2}^{(t)}$ are the cosine and sine components of $\mathbf{p}_n^{(t)}$, respectively. At the output of the final layer $t = J$, the continuous phasors are projected onto the nearest element of $\mathcal{F}$ and mapped to discrete phase indices via the same index-to-value mapping as the AO algorithm.

\subsubsection{Conservative Gain Assignment}

To ensure meaningful signal amplification across all considered values of $N_a$, the active-element gains are assigned a conservative discrete level that scales proportionally with the active-element density. The gain target is
\begin{equation}
\label{eq:gamma_c}
    \gamma_c \;=\; 1 + \beta\,\frac{N_a}{N_r},
\end{equation}
where $\beta > 0$ is selected to satisfy $\gamma_c \le \gamma_{\max}$, i.e., $\beta \le (\gamma_{\max}-1)\,N_r/N_a$, so that the target lies
within the valid amplification range $[1, \gamma_{\max}]$. The nearest discrete gain level in $\mathcal{G}$ is identified by
\begin{equation}
\label{eq:b_c}
    b_n^c \;=\;
    \underset{b\,\in\,\mathcal{B}}{\arg\min}\;
    \bigl|\mathcal{G}[b] - \gamma_c\bigr|,
    \quad \forall\, n \in \mathcal{A},
\end{equation}
and the fixed gain assigned to each element is
\begin{equation}
\label{eq:fixed_gains}
    \gamma_n^{(0)} \;=\;
    \begin{cases}
        \mathcal{G}[b_n^c], & n \in \mathcal{A}, \\[2pt]
        1,                   & n \in \mathcal{P}.
    \end{cases}
\end{equation}
Since $b_n^c \in \mathcal{B}$ and $\mathcal{G}[b_n^c] \in \mathcal{G}\subset [1,\gamma_{\max}]$ by construction, the assigned gain $\gamma_n^{(0)}$ is always a valid element of the discrete gain set, regardless of the specific value of $\gamma_c$. The corresponding fixed diagonal gain matrix is
$\boldsymbol{\Gamma}^{(0)} =
\mathrm{diag}(\gamma_1^{(0)},\ldots,\gamma_{N_r}^{(0)})
\in \mathbb{R}^{N_r \times N_r}$.
Should the total RF output power
$P_{\mathrm{RIS}} =
\sum_{n\in\mathcal{A}}(\gamma_n^{(0)})^2(P_{\mathrm{in},n}+\sigma_a^2)$
exceed $P_{\mathrm{RIS}}^{\max}$, the gain of the violating active element is reduced to the next lower discrete level and the budget re-evaluated until C3 is satisfied. Since $N_a \ll N_r$, this projection contributes negligible additional computational overhead.

\subsubsection{Unrolled Riemannian Gradient Ascent}

The DU network unrolls $J$ iterations of Riemannian gradient ascent on the sum-rate objective into a fixed chain of $J$ computational layers. Each layer $t \in \{1,\ldots,J\}$ applies two operations in sequence.

\paragraph{Gradient ascent step.}
The phasor matrix is updated in the direction of steepest sum-rate ascent:
\begin{equation}
\label{eq:gradient_step}
    \widetilde{\mathbf{P}}^{(t)} \;=\; \mathbf{P}^{(t-1)}
    \;+\; \alpha^{(t)}\,
    \nabla_{\mathbf{P}}\,R_{\mathrm{sum}}\!\left(\mathbf{P}^{(t-1)}\right),
\end{equation}
where $\nabla_{\mathbf{P}}\,R_{\mathrm{sum}}(\mathbf{P}^{(t-1)}) \in \mathbb{R}^{N_r \times 2}$ is the Euclidean gradient of the sum-rate with respect to the phasor state matrix, and $\alpha^{(t)} > 0$ is the learned step size at layer $t$.

\paragraph{Retraction to the unit circle.}
The gradient step displaces each row of $\widetilde{\mathbf{P}}^{(t)}$
off the unit circle. Each row is renormalized to restore unit length:
\begin{equation}
\label{eq:normalisation}
    \mathbf{p}_{n}^{(t)} \;=\;
    \frac{\tilde{\mathbf{p}}_{n}^{(t)}}
         {\|\tilde{\mathbf{p}}_{n}^{(t)}\|_2},
    \quad n = 1, \ldots, N_r,
\end{equation}
where $\tilde{\mathbf{p}}_{n}^{(t)}$ is the $n$-th row of
$\widetilde{\mathbf{P}}^{(t)}$. Equations~\eqref{eq:gradient_step}
and~\eqref{eq:normalisation} together constitute one Riemannian gradient ascent step on the product manifold $(\mathbb{S}^1)^{N_r}$.

\paragraph{Initialization.}
The unrolling begins from the identity phase configuration, consistent with the AO initialization:
\begin{equation}
\label{eq:init}
    \mathbf{p}_{n}^{(0)} \;=\; [1,\;0]^{\mathrm{T}},
    \quad n = 1, \ldots, N_r
    \quad(\theta_n^{(0)} = 0,\; q_n = 0).
\end{equation}

\paragraph{Learnable parameters.}
Each step size is parameterized as $\alpha^{(t)} = \exp(\tilde{\alpha}^{(t)})$ with $\tilde{\alpha}^{(t)} \in \mathbb{R}$ unconstrained, ensuring strict positivity. The complete set of trainable parameters is
\begin{equation}
\label{eq:alpha}
    \boldsymbol{\alpha} \;=\;
    \bigl[\alpha^{(1)},\, \ldots,\, \alpha^{(J)}\bigr]^{\mathrm{T}}
    \in \mathbb{R}_{>0}^{J}.
\end{equation}
The total number of trainable parameters is $J$, independent of $N_r$, $N_a$, $M$, and $K$.

\subsubsection{Differentiable Sum-Rate Evaluation}

The gradient $\nabla_{\mathbf{P}}\,R_{\mathrm{sum}}(\mathbf{P}^{(t-1)})$ in~\eqref{eq:gradient_step} is obtained by differentiating the sum-rate through the full physical signal chain. Given $\mathbf{P}^{(t-1)}$ and the fixed gain matrix $\boldsymbol{\Gamma}^{(0)}$, the complex phase-shift matrix at step $t-1$ is
\begin{equation}
\label{eq:Theta_j}
    \boldsymbol{\Theta}^{(t-1)} \;=\;
    \mathrm{diag}\!\left(v_1^{(t-1)},\, \ldots,\, v_{N_r}^{(t-1)}\right),
\end{equation}
with $v_n^{(t-1)}$ given by~\eqref{eq:complex_phasor}. The XL-RIS reflection matrix at step $t-1$ is
\begin{equation}
\label{eq:Psi_t}
    \boldsymbol{\Psi}^{(t-1)} \;=\;
    \boldsymbol{\Gamma}^{(0)}\,\boldsymbol{\Theta}^{(t-1)},
\end{equation}
and the effective cascaded channel is
\begin{equation}
\label{eq:Heff_t}
    \mathbf{H}_{\mathrm{eff}}^{(t-1)} \;=\;
    \mathbf{G}\,\boldsymbol{\Psi}^{(t-1)}\,\mathbf{H}
    \;\in\; \mathbb{C}^{K \times M}.
\end{equation}
The per-user effective noise variance under the fixed
gains~\eqref{eq:fixed_gains} is
\begin{equation}
\label{eq:noise_DU}
    \tilde{\sigma}_k^2 \;=\; \sigma_k^2
    \;+\; \sigma_a^2 \sum_{n \in \mathcal{A}}
    \!\left(\gamma_n^{(0)}\right)^{\!2} |g_{k,n}|^2,
    \quad k = 1, \ldots, K.
\end{equation}
Since the gains are fixed, $\tilde{\sigma}_k^2$ and the effective noise covariance matrix $\mathbf{R} = \mathrm{diag}(\tilde{\sigma}_1^2,\ldots,
\tilde{\sigma}_K^2)$ are constant across all $J$ layers and are computed once before the unrolling begins. The noise-aware MMSE precoder is updated from $\mathbf{H}_{\mathrm{eff}}^{(t-1)}$ and $\mathbf{R}$ via
\eqref{eq:mmse_user_domain}, scaled to satisfy C2, and the sum-rate is evaluated via~\eqref{eq:SINR} and~\eqref{eq:sum_rate}. Since every operation in this chain is differentiable with respect to $\mathbf{P}^{(t-1)}$, the gradient $\nabla_{\mathbf{P}}\,R_{\mathrm{sum}}(\mathbf{P}^{(t-1)})$ is computed exactly via automatic differentiation through the entire signal chain.

\subsubsection{Self-Supervised Training}

The step sizes $\boldsymbol{\alpha}$ are optimized to maximize the sum-rate of the quantized output over a training set $\{(\mathbf{H}^{(i)},\mathbf{G}^{(i)})\}_{i=1}^{N_{\mathrm{train}}}$ of raw channel realizations. No labeled data are required; the sole supervisory signal is the sum-rate itself.

\paragraph{Straight-through estimator.}
The argmax projection that maps $\mathbf{P}^{(J)}$ to discrete phase indices is non-differentiable. A STE~\cite{bengio2013ste} is applied to propagate gradients through this step. In the \emph{forward pass}, the discrete phase index for element $n$ is selected as
\begin{equation}
\label{eq:STE_forward}
    \hat{q}_n \;=\;
    \underset{q\,\in\,\mathcal{Q}}{\arg\max}\;
    \mathbf{p}_{n}^{(J)} \cdot
    \!\left[\cos\tfrac{2\pi q}{L_\theta},\;
             \sin\tfrac{2\pi q}{L_\theta}\right]^{\mathrm{T}},
\end{equation}
and the quantized phasor $\hat{\mathbf{p}}_n =
[\cos(2\pi\hat{q}_n/L_\theta),\;\sin(2\pi\hat{q}_n/L_\theta)]^{\mathrm{T}}$ is used for sum-rate evaluation. In the \emph{backward pass}, the gradient is propagated through the quantization step as if it were the identity mapping, i.e., $\partial\hat{\mathbf{p}}_n/\partial\mathbf{p}_n^{(J)} \approx I_2$.
This approximation is well-suited to the present setting for two reasons. First, after $J$ gradient ascent steps, each phasor $\mathbf{p}_n^{(J)}$ lies close to a codebook direction, so the quantization error is small and the identity approximation is locally accurate. Second, unlike
Gumbel-softmax relaxations, the STE applies the exact argmax in the forward pass, ensuring that the training objective matches the inference objective exactly and eliminating any temperature-induced bias.

\paragraph{Training loss.}
The training objective minimized over $\boldsymbol{\alpha}$ is
\begin{equation}
\label{eq:loss}
    \mathcal{L}(\boldsymbol{\alpha}) \;=\;
    -\frac{1}{|\mathcal{I}|} \sum_{i \in \mathcal{I}}
    R_{\mathrm{sum}}^{\mathrm{STE}}\!\left(\mathbf{P}_i^{(J)}\right),
\end{equation}
where $\mathcal{I}$ is a mini-batch of sample indices and
$R_{\mathrm{sum}}^{\mathrm{STE}}(\mathbf{P}_i^{(J)})$ denotes the sum-rate evaluated using the quantized phasors~\eqref{eq:STE_forward} in the forward pass with continuous-phasor gradients in the backward pass.
The parameters $\boldsymbol{\alpha}$ are updated via the AdamW optimizer~\cite{loshchilov2019decoupled} with gradient clipping at unit norm and a cosine learning-rate schedule over $N_{\mathrm{ep}}$ epochs.

\subsubsection{Inference and Discrete Configuration Recovery}
\label{sec:DU_inference}

At deployment, given a new channel realization $(\mathbf{H},\mathbf{G})$, the trained DU network runs $J$ gradient ascent steps and recovers the discrete XL-RIS configuration as summarized in Algorithm~\ref{alg:DU}.

\paragraph{Phase index recovery.}
After $J$ steps with the trained $\boldsymbol{\alpha}$, the discrete phase index of each element $n$ is
\begin{equation}
\label{eq:quantise}
    q_n^* \;=\;
    \underset{q\,\in\,\mathcal{Q}}{\arg\max}\;
    \mathbf{p}_{n}^{(J)} \cdot
    \!\left[\cos\tfrac{2\pi q}{L_\theta},\;
             \sin\tfrac{2\pi q}{L_\theta}\right]^{\mathrm{T}},
    \quad n = 1, \ldots, N_r,
\end{equation}
and the corresponding phase value is $\theta_n^* = 2\pi q_n^*/L_\theta \in \mathcal{F}$.

\paragraph{Gain index recovery.}
Each active element is assigned the conservative gain computed in~\eqref{eq:fixed_gains}, subject to the downward projection described above. The recovered gain index and value are
\begin{equation}
\label{eq:gain_recovery}
    b_n^* \;=\; b_n^c,
    \quad
    \gamma_n^* \;=\; \mathcal{G}[b_n^*],
    \quad \forall\, n \in \mathcal{A},
\end{equation}
and passive elements retain unit gain: $\gamma_n^* = 1$ for all $n \in \mathcal{P}$.

\paragraph{XL-RIS configuration and BS precoder.}
The recovered discrete indices determine
\begin{equation}
\label{eq:Psi_hat}
    \hat{\boldsymbol{\Psi}} \;=\;
    \hat{\boldsymbol{\Gamma}}\hat{\boldsymbol{\Theta}} \;=\;
    \mathrm{diag}\!\left(
    \gamma_1^* e^{j\theta_1^*},\ldots,\gamma_{N_r}^* e^{j\theta_{N_r}^*}
    \right),
\end{equation}
where $\hat{\boldsymbol{\Theta}} =
\mathrm{diag}(e^{j\theta_1^*},\ldots,e^{j\theta_{N_r}^*})$ and
$\hat{\boldsymbol{\Gamma}} =
\mathrm{diag}(\gamma_1^*,\ldots,\gamma_{N_r}^*)$.
Given $\hat{\boldsymbol{\Psi}}$, the effective channel
$\mathbf{H}_{\mathrm{eff}} = \mathbf{G}\hat{\boldsymbol{\Psi}}\mathbf{H}$
is formed and the final BS precoder is computed
via~\eqref{eq:mmse_user_domain}.

\begin{algorithm}[t]
\small
\caption{DU-Based XL-RIS Phase Configuration: Inference}
\label{alg:DU}
\begin{algorithmic}[1]
\REQUIRE $\mathbf{H},\,\mathbf{G}$;\;
         trained $\boldsymbol{\alpha}=[\alpha^{(1)},\ldots,\alpha^{(J)}]^{\mathrm{T}}$;\;
         $\boldsymbol{\Gamma}^{(0)}$, $\mathcal{Q}$,
         $P_{\mathrm{TX}}$, $\sigma_k^2$, $\sigma_a^2$, $\beta$
\ENSURE  $\mathbf{q}^*$;\;
         $\mathbf{b}^*$;\;
         $\hat{\boldsymbol{\Psi}}$;\;
         $\mathbf{W}$
\STATE Compute $\gamma_c \leftarrow 1 + \beta\,(N_a/N_r)$;\;
       $b_n^c \leftarrow \arg\min_{b\in\mathcal{B}}\,|\mathcal{G}[b]-\gamma_c|$
       for $n \in \mathcal{A}$
       \COMMENT{Conservative gain target via~\eqref{eq:gamma_c}}
\STATE Set $\gamma_n^{(0)} \leftarrow \mathcal{G}[b_n^c]$ for
       $n\in\mathcal{A}$;\;
       $\gamma_n^{(0)} \leftarrow 1$ for $n\in\mathcal{P}$;\;
       project downward if $P_{\mathrm{RIS}} > P_{\mathrm{RIS}}^{\max}$
       \COMMENT{via~\eqref{eq:fixed_gains}}
\STATE $\tilde{\sigma}_k^2 \leftarrow
       \sigma_k^2 + \sigma_a^2\sum_{n\in\mathcal{A}}
       (\gamma_n^{(0)})^2|g_{k,n}|^2$,\;
       $k=1,\ldots,K$;\;
       $\mathbf{R} \leftarrow
       \mathrm{diag}(\tilde{\sigma}_1^2,\ldots,\tilde{\sigma}_K^2)$
       \COMMENT{Constant across all $J$ layers}
\STATE $\mathbf{p}_n^{(0)} \leftarrow [1,\,0]^{\mathrm{T}}$
       for all $n$
       \COMMENT{Identity initialization: $\theta_n^{(0)}=0$}
\FOR{$t=1$ \TO $J$}
    \STATE $v_n^{(t-1)} \leftarrow p_{n,1}^{(t-1)}+j\,p_{n,2}^{(t-1)}$
           for all $n$
           \COMMENT{via~\eqref{eq:complex_phasor}}
    \STATE $\boldsymbol{\Theta}^{(t-1)} \leftarrow
           \mathrm{diag}(v_1^{(t-1)},\ldots,v_{N_r}^{(t-1)})$;\;
           $\boldsymbol{\Psi}^{(t-1)} \leftarrow
           \boldsymbol{\Gamma}^{(0)}\boldsymbol{\Theta}^{(t-1)}$
    \STATE $\mathbf{H}_{\mathrm{eff}}^{(t-1)} \leftarrow
           \mathbf{G}\,\boldsymbol{\Psi}^{(t-1)}\,\mathbf{H}$
           \COMMENT{via~\eqref{eq:Heff_t}}
    \STATE Compute $\mathbf{W}^{(t-1)}$ via~\eqref{eq:mmse_user_domain};
           scale to $\|\mathbf{W}^{(t-1)}\|_F^2 = P_{\mathrm{TX}}$
    \STATE Evaluate $R_{\mathrm{sum}}(\mathbf{P}^{(t-1)})$
           via~\eqref{eq:SINR}--\eqref{eq:sum_rate}
    \STATE $\nabla_{\mathbf{P}}\,R_{\mathrm{sum}}(\mathbf{P}^{(t-1)})
           \leftarrow$ using automatic differentiation
    \STATE $\widetilde{\mathbf{P}}^{(t)} \leftarrow \mathbf{P}^{(t-1)}
           + \alpha^{(t)}\,\nabla_{\mathbf{P}}\,
           R_{\mathrm{sum}}(\mathbf{P}^{(t-1)})$
           \COMMENT{via~\eqref{eq:gradient_step}}
    \STATE $\mathbf{p}_n^{(t)} \leftarrow
           \tilde{\mathbf{p}}_n^{(t)}/\|\tilde{\mathbf{p}}_n^{(t)}\|_2$
           for all $n$
           \COMMENT{Retraction via~\eqref{eq:normalisation}}
\ENDFOR
\FOR{$n=1$ \TO $N_r$}
    \STATE $q_n^* \leftarrow
           \arg\max_{q\in\mathcal{Q}}\;
           \mathbf{p}_n^{(J)}\cdot
           [\cos(2\pi q/L_\theta),\;\sin(2\pi q/L_\theta)]^{\mathrm{T}}$
           \COMMENT{via~\eqref{eq:quantise}}
    \STATE $\theta_n^* \leftarrow 2\pi q_n^*/L_\theta \in \mathcal{F}$
\ENDFOR
\STATE $b_n^* \leftarrow b_n^c$,\;
       $\gamma_n^* \leftarrow \mathcal{G}[b_n^c]$
       for $n\in\mathcal{A}$;\;
       $\gamma_n^*\leftarrow 1$ for $n\in\mathcal{P}$
       \COMMENT{via~\eqref{eq:gain_recovery}}
\STATE $\hat{\boldsymbol{\Psi}} \leftarrow
       \mathrm{diag}(\gamma_n^*e^{j\theta_n^*})_{n=1}^{N_r}$
       \COMMENT{via~\eqref{eq:Psi_hat}}
\STATE $\mathbf{H}_{\mathrm{eff}} \leftarrow
       \mathbf{G}\hat{\boldsymbol{\Psi}}\mathbf{H}$;\;
       compute $\mathbf{W}$ via~\eqref{eq:mmse_user_domain},
       scale to $\|\mathbf{W}\|_F^2=P_{\mathrm{TX}}$
\RETURN $\mathbf{q}^*,\;\mathbf{b}^*,\;\hat{\boldsymbol{\Psi}},\;\mathbf{W}$
\end{algorithmic}
\end{algorithm}

\subsection{MO Baseline}
\label{subsec:mo_baseline}

To provide a continuous-domain benchmark, an MO scheme is included for comparison. The XL-RIS phase coefficients are optimized over the unit-modulus manifold $(\mathbb{S}^1)^{N_r}$, i.e., each coefficient satisfies $|\psi_n| = 1$, using Riemannian gradient ascent iterations.
The active-element gains are fixed at the same conservative level $\boldsymbol{\Gamma}^{(0)}$ defined in~\eqref{eq:fixed_gains}, subject to the same RF power projection as the DU framework. Upon convergence, the continuous solution is quantized to the nearest discrete phase level in $\mathcal{F}$ to ensure hardware compatibility.

\section{Complexity Analysis}
\label{sec:complexity_analysis}

This section analyzes the computational complexity of the AO algorithm, the MO baseline, and the proposed DU framework. The objective is to characterize the scaling behavior with respect to the number of XL-RIS elements $N_r$, active elements $N_a$, and users $K$. The AO
implementation employs local neighborhood search and multiple random restarts; therefore, the complexity expressions reflect this structure rather than exhaustive enumeration of the discrete codebooks.

\subsection{Complexity of the AO Algorithm}

The AO procedure alternates between BS precoder updates and discrete coordinate-wise refinement of the XL-RIS phase and gain indices. Let $T_{\mathrm{AO}}$ denote the number of iterations per restart and $R$ the number of independent restarts.

\subsubsection{BS Precoder Update}

For a fixed XL-RIS configuration $\boldsymbol{\Psi}$, the effective cascaded channel $\mathbf{H}_{\mathrm{eff}} = \mathbf{G}\boldsymbol{\Psi}
\mathbf{H} \in \mathbb{C}^{K \times M}$ in~\eqref{eq:Heff} is constructed with complexity $\mathcal{O}(K N_r M)$. The noise-aware MMSE update in~\eqref{eq:mmse_user_domain} subsequently requires: (i)~forming the
$K \times K$ Gram matrix
$\mathbf{H}_{\mathrm{eff}}\mathbf{H}_{\mathrm{eff}}^{\mathrm{H}}$ at cost $\mathcal{O}(K^2 M)$, and (ii)~inverting the resulting $K \times K$ system at cost $\mathcal{O}(K^3)$. One complete precoder update therefore incurs a complexity of $\mathcal{O}(K N_r M + K^2 M + K^3)$.
In the XL-RIS regime where $N_r \gg K$, the dominant term reduces to $\mathcal{O}(K N_r M)$.

\subsubsection{Discrete Phase Optimization}

The phase indices are updated via discrete coordinate ascent as described in~\eqref{eq:phase_update}. For each element $n \in \{1,\ldots,N_r\}$, the neighborhood $\mathcal{N}_{\mathcal{Q}}(q_n)$ in~\eqref{eq:phase_nbhd} contains exactly two candidates, independently of $L_\theta$. Evaluating each candidate does \emph{not} require reconstructing $\mathbf{H}_{\mathrm{eff}}$ from scratch: since only the $n$-th element is perturbed, the change in $\mathbf{H}_{\mathrm{eff}}$ is a rank-one update confined to its $n$-th column, which can be re-evaluated at cost $\mathcal{O}(KM)$ per candidate. A complete sweep over all $N_r$ phase indices therefore incurs $\mathcal{O}(KMN_r)$, which is linear in $N_r$,
in contrast to the quadratic cost of full channel reconstruction at every coordinate update.

\subsubsection{Discrete Active-Gain Optimization}

Each active element $n \in \mathcal{A}$ performs an analogous constant-size neighborhood search over
$\mathcal{N}_{\mathcal{B}}(b_n)$ in~\eqref{eq:gain_nbhd}, with cardinality at most two. By the same rank-one update argument, each candidate evaluation costs $\mathcal{O}(KM)$, and updating all $N_a$ active gain indices in one sweep requires $\mathcal{O}(N_a KM)$.
Since $N_a \le N_r$, this term is of equal or lower order than the phase refinement sweep.

\subsubsection{Overall AO Complexity}

The per-iteration complexity within one restart is dominated by the phase sweep, yielding $\mathcal{O}(KMN_r)$. Accounting for the precoder update, which is also $\mathcal{O}(KN_rM)$, the total cost per iteration remains $\mathcal{O}(KMN_r)$. Over $T_{\mathrm{AO}}$ iterations and $R$ independent restarts, the overall AO complexity is $\mathcal{O}(R\,T_{\mathrm{AO}}\,K\,M\,N_r)$.
The AO algorithm therefore exhibits \emph{linear} scaling in $N_r$, as the local neighborhood structure limits each coordinate update to a constant number of rank-one evaluations, with multiple restarts increasing the total cost linearly in $R$.

\subsection{Complexity of the MO Benchmark}

In the MO benchmark, the XL-RIS reflection vector is optimized over the unit-modulus manifold $(\mathbb{S}^1)^{N_r}$. Each Riemannian
gradient iteration requires evaluating the objective and its gradient with respect to the $N_r$ reflection coefficients, incurring a complexity of $\mathcal{O}(KN_rM)$. Retraction and projection operations contribute an additional $\mathcal{O}(N_r)$ cost, which is of lower order. Let $T_{\mathrm{MO}}$ denote the number of MO iterations. The overall complexity is $\mathcal{O}(T_{\mathrm{MO}}\,K\,N_r\,M)$.
Although MO avoids discrete combinatorial search, it remains iterative and must be executed independently for each channel realization, resulting in substantial computational overhead as $N_r$ increases.

\subsection{Complexity of the Proposed DU Framework}

\subsubsection{Training Complexity}

During offline training, the DU framework unfolds $J$ gradient-update layers for each channel realization. At every layer, one evaluation of the effective channel, the MMSE precoder, the SINRs, and the sum-rate gradient is performed via automatic differentiation. The dominant cost at each layer is the effective channel construction $\mathbf{H}_{\mathrm{eff}} =
\mathbf{G}\boldsymbol{\Psi}\mathbf{H}$, scaling as $\mathcal{O}(KN_rM)$, so the overall training complexity per channel realization is $\mathcal{O}(J\,K\,M\,N_r)$. Since training is performed entirely offline, this cost does not affect real-time deployment.

\subsubsection{Inference Complexity}

At deployment, the DU framework executes exactly $J$ gradient-update steps per channel realization, each incurring the same dominant cost, giving an inference complexity of $\mathcal{O}(J\,K\,M\,N_r)$. Since $J$ is a small fixed constant, this expression is independent of the number of restarts or convergence iterations. Within the system dimensions considered ($N_r \le 512$, $K \le 16$), the measured inference runtime remains nearly
constant across both $N_r$ and $K$: the fixed per-step overhead of automatic differentiation dominates over the arithmetic scaling at these scales, so the theoretical linear dependence on $N_r$ does not manifest in practice within the considered range.

\subsection{Complexity Comparison}

Table~\ref{tab:complexity} summarizes the computational complexity of the considered schemes. Both the AO algorithm and the MO benchmark require multiple iterative evaluations per channel realization, resulting in higher cost as the number of iterations grows. In contrast, the DU framework employs a fixed number of unrolled gradient-update layers,
yielding significantly lower runtime complexity while preserving linear scaling with $N_r$.

\begin{table}[!t]
    \centering
    \caption{Computational Complexity Comparison}
    \label{tab:complexity}
    \begin{tabular}{lc}
        \toprule
        \textbf{Scheme} & \textbf{Complexity} \\
        \midrule
        AO
        & $\mathcal{O}(R\,T_{\mathrm{AO}}\,K\,M\,N_r)$ \\[4pt]
        MO (Benchmark)
        & $\mathcal{O}(T_{\mathrm{MO}}\,K\,M\,N_r)$ \\[4pt]
        Proposed DU
        & $\mathcal{O}(J\,K\,M\,N_r)$ \\
        \bottomrule
    \end{tabular}
\end{table}

\section{Simulation Setup and Results}
\label{sec:simulation_setup}

This section evaluates the performance of the proposed hybrid XL-RIS beamforming framework under the near-field spherical-wave propagation model introduced in Section~\ref{sec:system_model}. All numerical results are averaged over a sufficiently large number of independently generated channel realizations. Unless otherwise stated, all schemes are evaluated under identical channel realizations, deployment geometry, and power constraints.

\subsection{Simulation Setup}

The default system parameters are summarized in Table~\ref{tab:sim_params}. The BS employs $M = 64$ transmit antennas to simultaneously serve $K = 8$ single-antenna users through a hybrid XL-RIS comprising $N_r = 256$ reflecting elements, of which $N_a = 16$ are active. The carrier frequency is $f_c = 3.5~\mathrm{GHz}$, corresponding to a wavelength $\lambda \approx 0.0857~\mathrm{m}$, and adjacent XL-RIS elements are spaced
by $\lambda/2$. The resulting aperture of approximately
$1.33~\mathrm{m} \times 0.30~\mathrm{m}$ yields a Rayleigh distance of $d_R \approx 41.2~\mathrm{m}$. User locations are randomly drawn from the region $(x,y,z) = (55,\;[-15,15],\;1.5)~\mathrm{m}$, which lies within the forward Fresnel region, so near-field spherical-wave propagation applies. Discrete phase control with $B = 3$ bits yields $L_\theta = 2^B = 8$ quantization levels. Active elements employ eight uniformly spaced gain levels within $[1, 2]$. The BS transmit power is $P_{\mathrm{TX}} = 50~\mathrm{dBm}$, receiver and amplifier noise variances are $\sigma_k^2 = \sigma_a^2 = 10^{-4}$, and the XL-RIS RF output power budget is $P_{\mathrm{RIS}}^{\max} = 42~\mathrm{dBm}$.

The DU framework is trained offline on randomly generated channel realizations. The network unrolls $J = 6$ Riemannian gradient-update layers, each parameterized by a single learnable positive step size. Active-element gains are assigned the conservative discrete level $\gamma_c = 1 + \beta N_a/N_r$ nearest in $\mathcal{G}$, with downward projection applied only if the RF power budget is violated.
At inference, the discrete XL-RIS phase configuration is recovered via quantized phasor projection as described in Section~\ref{sec:DU_inference}.

The following five schemes are evaluated:
\begin{itemize}
    \item \textbf{AO}: the proposed DCA-based alternating optimization algorithm serving as the near-optimal scheme.
    \item \textbf{DU + Ref}: the proposed DU framework followed by one additional global $\pm 1$ discrete phase coordinate sweep with active-element gains held fixed.
    \item \textbf{DU}: the proposed deep unfolding framework operating without any post-inference refinement.
    \item \textbf{MO}: manifold optimization over the continuous unit-modulus manifold $(\mathbb{S}^1)^{N_r}$, with active-element gains fixed at the conservative level, quantized to $\mathcal{F}$ upon convergence.
    \item \textbf{Random}: randomly drawn discrete phase and gain configurations, serving as a lower-bound reference.
\end{itemize}

\begin{table}[!t]
\centering
    \caption{Default Simulation and DU Training Parameters}
    \label{tab:sim_params}
    \small
    \renewcommand{\arraystretch}{1.05}
    \newcolumntype{R}{>{\raggedleft\arraybackslash}X}
    \begin{tabularx}{\linewidth}{@{}p{0.52\linewidth}>{\arraybackslash}X@{}}
    
    \toprule
    \multicolumn{2}{c}{\textbf{System Simulation Parameters}} \\
    \midrule
    Carrier frequency ($f_c$)                    & $3.5~\mathrm{GHz}$                                        \\
    Wavelength ($\lambda$)                       & ${\approx}0.0857~\mathrm{m}$                              \\
    BS antennas ($M$), users ($K$)               & $M = 64,\; K = 8$                                         \\
    Total XL-RIS elements ($N_r$)                & $256$                                                     \\
    Active XL-RIS elements ($N_a$)               & $16$                                                      \\
    XL-RIS element spacing                       & ${\approx}0.0429~\mathrm{m}$                              \\
    Phase resolution ($B$)                       & $3~\mathrm{bits}$                                         \\
    Phase levels ($L_\theta$)                    & $2^B = 8$                                                 \\
    Active gain levels ($|\mathcal{G}|$)         & $8$                                                       \\
    BS transmit power ($P_{\mathrm{TX}}$)        & $50~\mathrm{dBm}$                                         \\
    Receiver noise variance ($\sigma_k^2$)       & $10^{-4}$                                                 \\
    Amplifier noise variance ($\sigma_a^2$)      & $10^{-4}$                                                 \\
    XL-RIS RF budget ($P_{\mathrm{RIS}}^{\max}$) & $42~\mathrm{dBm}$                                         \\
    BS location                                  & $(0,0,10)~\mathrm{m}$                                     \\
    XL-RIS centre location                       & $(40,0,6)~\mathrm{m}$                                     \\
    UE region                                    & $(55,[-15,15],1.5)~\mathrm{m}$                            \\
    BS--XL-RIS distance ($d_{\mathrm{BR}}$)      & $40~\mathrm{m}$                                           \\
    XL-RIS--UE distance ($d_{\mathrm{RU}}$)      & $15~\mathrm{m}$                                           \\
    XL-RIS aperture                              & $W{\approx}1.33~\mathrm{m}$, $H{\approx}0.30~\mathrm{m}$ \\
    Rayleigh distance ($d_R$)                    & ${\approx}41.2~\mathrm{m}$                 \\
    
    \midrule
    \multicolumn{2}{c}{\textbf{DU Training Configuration}} \\
    \midrule
    Learning paradigm                            & Self-supervised                                            \\
    Unfolding layers ($J$)                       & $6$                                                        \\
    Trainable parameters                         & $J$ step sizes $\{\alpha^{(t)}\}_{t=1}^{J}$ 
                                                \\
    Optimizer                                    & AdamW                                                      \\
    Initial learning rate                        & $10^{-3}$                                                  \\
    Weight decay                                 & $5\times10^{-5}$                                           \\
    Gradient clipping norm                       & $1$                                                        \\
    LR schedule                                  & Cosine annealing                                           \\
    Batch size                                   & $16$                                                       \\
    Training epochs                              & $350$                                                      \\
    Training samples                             & $1000$                                                     \\
    Training objective                           & Negative sum-rate (STE)                                    \\
    \bottomrule
    \end{tabularx}
\end{table}
\subsection{Results and Discussion}

\subsubsection{Achievable Sum-Rate versus SNR}

\begin{figure}
\centering
    \includegraphics[width=0.95\columnwidth]{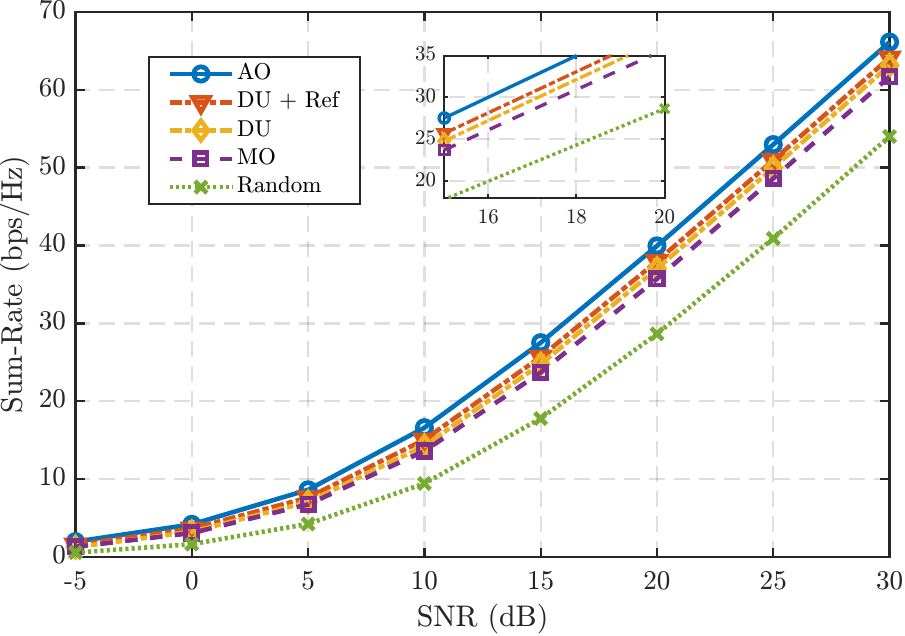}
    \caption{Achievable sum-rate versus transmit SNR for $N_r=256$, $M=64$, $K=8$, $P_{\mathrm{TX}}=50~\mathrm{dBm}$, and $P_{\mathrm{RIS}}^{\max}=42~\mathrm{dBm}$.}
    \label{fig:sumrate}
\end{figure}

Fig.~\ref{fig:sumrate} shows the achievable sum-rate as a function of transmit SNR, varied from $-5~\mathrm{dB}$ to $30~\mathrm{dB}$. The sum-rate increases monotonically with SNR for all schemes.
The AO algorithm achieves the highest sum-rate across the entire range, as its iterative coordinate-ascent updates jointly optimize the discrete phase shifts, active-element gains, and BS precoder under the RF power constraint. 
DU + Ref closely follows AO, with the post-inference refinement sweep recovering part of the residual gap relative to the unrefined DU. DU itself maintains a small but consistent gap below DU + Ref throughout the SNR range.
MO ranks below DU: its continuous unit-modulus optimization yields a competitive solution in the unconstrained domain, but discrete phase quantization and RF power constraint enforcement degrade its performance relative to the DU schemes. Random provides the lower bound.
\subsubsection{Impact of XL-RIS Size}

\begin{figure}
\centering
    \includegraphics[width=0.95\columnwidth]{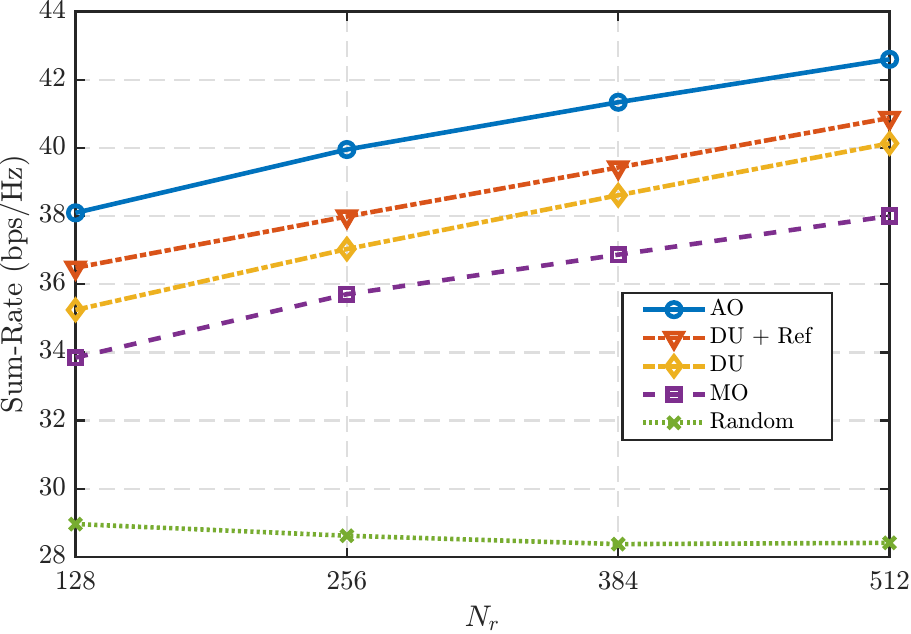}
    \caption{Achievable sum-rate versus the number of XL-RIS elements for $M=64$, $K=8$, $\mathrm{SNR}=20~\mathrm{dB}$, $P_{\mathrm{TX}}=50~\mathrm{dBm}$, and $P_{\mathrm{RIS}}^{\max}=42~\mathrm{dBm}$.}
    \label{fig:Nr_sr}
\end{figure}

\begin{figure}
\centering
    \includegraphics[width=0.95\columnwidth]{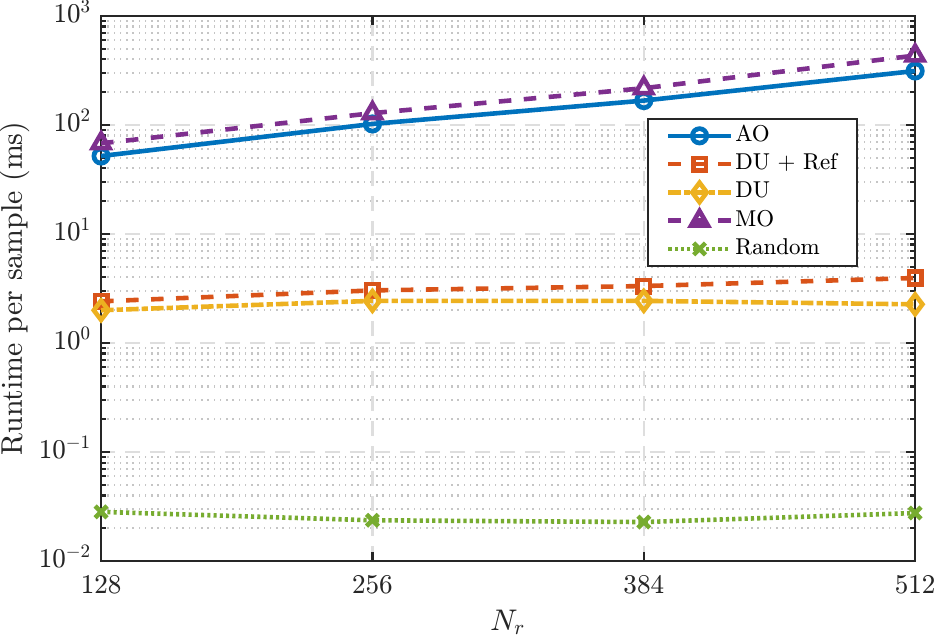}
    \caption{Runtime per channel realization versus the number of XL-RIS elements for $M=64$, $K=8$, and $\mathrm{SNR}=20~\mathrm{dB}$.}
    \label{fig:Nr_runtime}
\end{figure}

Fig.~\ref{fig:Nr_sr} shows the achievable sum-rate as $N_r$ increases from 128 to 512. All schemes improve with $N_r$ as the larger aperture enhances near-field spatial focusing and effective aperture gain. The sum-rate ordering AO, DU\,+\,Ref, DU, MO, and Random is maintained across all considered array sizes. The gap between AO and DU remains approximately constant, indicating that the DU generalizes well as $N_r$ grows.
Fig.~\ref{fig:Nr_runtime} shows the corresponding runtime per channel realization. DU and DU + Ref are the fastest and exhibit nearly constant runtime across the entire $N_r$ range and remain almost indistinguishable from each other throughout: the $J = 6$ autodiff-based gradient evaluations take a fixed cost that is dominated by framework overhead rather than the matrix dimension, so the DU runtime does not increase noticeably as $N_r$ grows from 128 to 512. The single refinement sweep in DU + Ref adds one coordinate pass over $N_r$ elements, but this cost remains small relative to the autodiff overhead, keeping the two curves almost overlapping. AO runtime grows substantially with $N_r$ due to repeated channel evaluations and iterative discrete variable updates across $R$ restarts. MO exhibits the highest runtime among all schemes, exceeding AO: unlike AO, which terminates each restart as soon as no further improvement is observed, MO must iterate to continuous convergence before quantization, requiring a larger total number of gradient evaluations per channel realization.

\subsubsection{Impact of Active Element Count}

\begin{figure}
\centering
    \includegraphics[width=0.95\columnwidth]{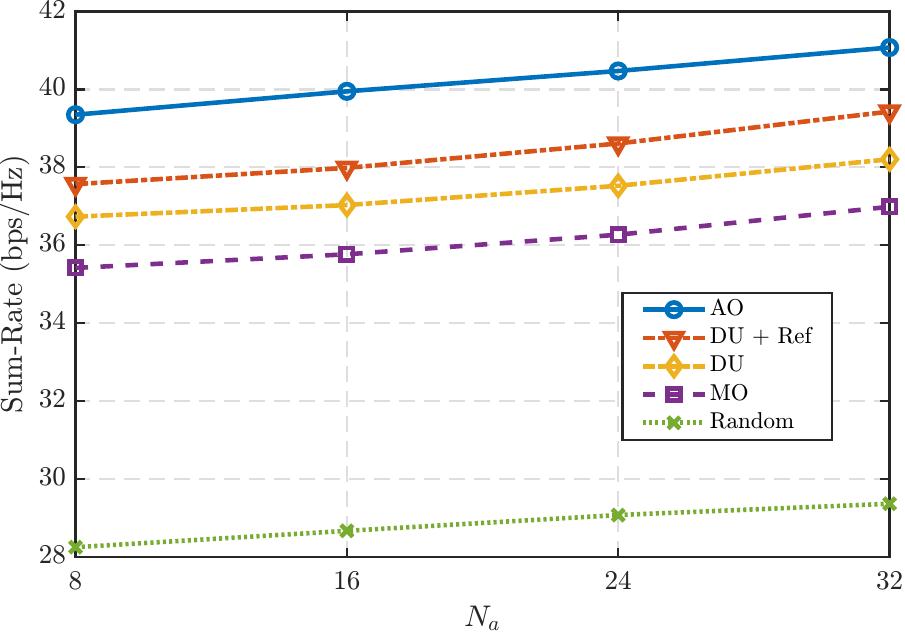}
    \caption{Achievable sum-rate versus the number of active XL-RIS elements for $N_r=256$, $M=64$, $K=8$, $\mathrm{SNR}=20~\mathrm{dB}$,
    $P_{\mathrm{TX}}=50~\mathrm{dBm}$, and
    $P_{\mathrm{RIS}}^{\max}=42~\mathrm{dBm}$.}
    \label{fig:Nactive_sr}
\end{figure}

Fig.~\ref{fig:Nactive_sr} shows the achievable sum-rate as $N_a$ varies from 8 to 32 with $N_r = 256$ fixed. Increasing $N_a$ raises the reflected signal power through controlled amplification. However, each additional active element also introduces amplifier noise and consumes part of the RF power
budget. Within the considered range, the amplification benefit outweighs the associated noise and power costs, so all schemes exhibit a moderate, consistent sum-rate improvement with $N_a$. The relative ordering AO, DU\,+\,Ref, DU, MO, and Random is preserved across all values of $N_a$.

\subsubsection{Impact of Number of Users}

\begin{figure}
\centering
    \includegraphics[width=0.95\columnwidth]{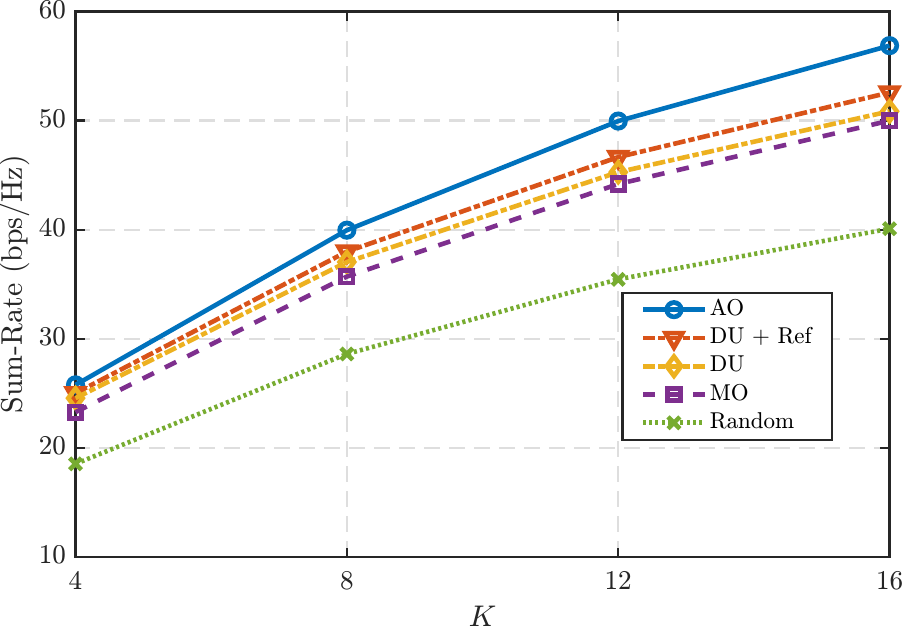}
    \caption{Achievable sum-rate versus the number of users for $N_r=256$, $M=64$, $\mathrm{SNR}=20~\mathrm{dB}$,
    $P_{\mathrm{TX}}=50~\mathrm{dBm}$, and
    $P_{\mathrm{RIS}}^{\max}=42~\mathrm{dBm}$.}
    \label{fig:K_sr}
\end{figure}

Fig.~\ref{fig:K_sr} shows the achievable sum-rate as $K$ increases from 4 to 16. The sum-rate rises with $K$ for all schemes as additional spatial data streams are simultaneously transmitted. The large BS antenna array provides sufficient spatial degrees of freedom to support multi-user transmission, and the hybrid XL-RIS improves effective channel quality through passive phase alignment and active amplification. The growth rate diminishes at higher $K$ as inter-user interference intensifies and the XL-RIS configuration must simultaneously serve more propagation paths. The ordering AO, DU\,+\,Ref, DU, MO, and Random is maintained throughout.
\begin{figure}
\centering
    \includegraphics[width=0.95\columnwidth]{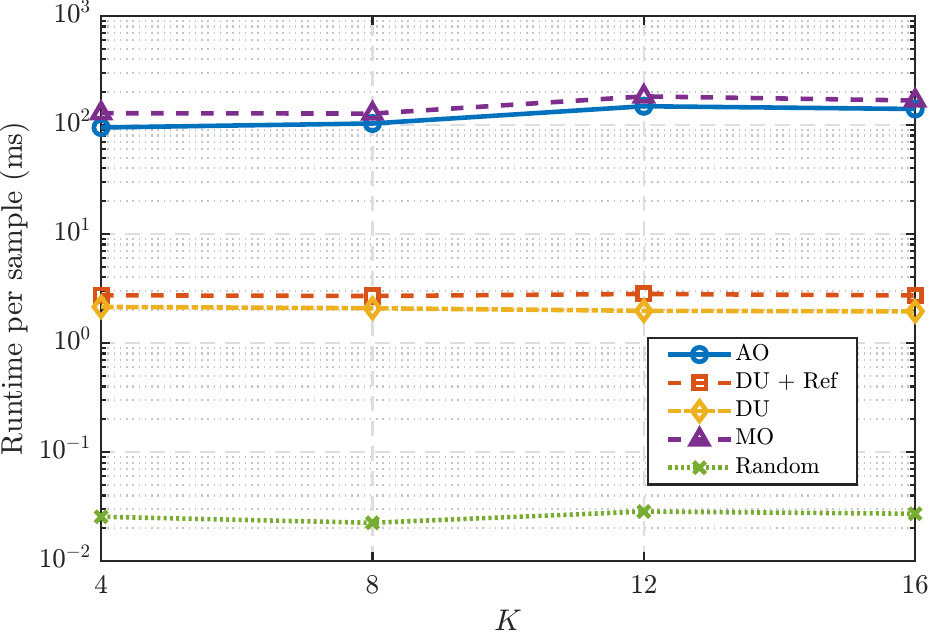}
    \caption{Runtime per channel realization versus the number of users for $N_r=256$, $M=64$, $\mathrm{SNR}=20~\mathrm{dB}$,
    $P_{\mathrm{TX}}=50~\mathrm{dBm}$, and
    $P_{\mathrm{RIS}}^{\max}=42~\mathrm{dBm}$.}
    \label{fig:K_runtime}
\end{figure}
Fig.~\ref{fig:K_runtime} shows the runtime as $K$ increases. DU and DU + Ref remain the fastest and exhibit nearly constant runtime across the entire $K$ range, and the two curves remain close throughout: for $K \in \{4,\ldots,16\}$, the tensor sizes
processed at each of the $J = 6$ gradient steps are small enough that the fixed autodiff overhead dominates over the arithmetic scaling with $K$, making the DU runtime effectively flat. The single refinement sweep in DU + Ref similarly contributes a cost that remains negligible at these scales, keeping the two curves overlapping. AO exhibits substantially higher runtime due to repeated multi-user channel evaluations and precoder updates across $R$ restarts. MO remains the slowest scheme: iterating to continuous convergence before quantization requires more gradient evaluations than the AO restart strategy, and the associated cost grows more steeply with $K$.

Across all evaluated configurations, the sum-rate ordering AO, DU\,+\,Ref, DU, MO, and Random is consistent, while the runtime ordering (fastest to slowest) is Random, DU, DU\,+\,Ref, AO, and MO. The DU and DU + Ref runtimes remain nearly constant across the $N_r$ and $K$ sweeps, confirming that the $J = 6$ fixed gradient evaluations incur a cost dominated by autodiff overhead rather than by arithmetic scaling with system dimensions within the considered ranges. MO is the most computationally demanding scheme because continuous manifold convergence requires more iterations than the AO restart strategy, yet its sum-rate falls below both DU and DU\,+\,Ref after discrete quantization and RF power constraint enforcement. These results confirm that the proposed DU framework provides a favorable and consistent trade-off: near-AO sum-rate performance with runtime substantially below AO and MO across all system configurations.

\section{Conclusion}
\label{sec:conclusion}

This paper investigated hybrid XL-RIS-assisted downlink MU-MISO transmission in the near-field regime under practical hardware constraints, including discrete phase control, quantized amplification gains, and a strict RF output power budget. The joint beamforming design was formulated as a mixed discrete--continuous sum-rate maximization problem and addressed through two complementary schemes.
A DCA-based AO framework was first developed as the
near-optimal scheme. The discrete XL-RIS phase and gain variables were updated via local index-domain searches over the respective codebooks, avoiding exhaustive enumeration, while the BS precoder was computed in closed form using a noise-aware MMSE criterion. Multiple random restarts were incorporated to improve robustness against local stationary points. A low-complexity DU framework was then proposed as an alternative for scalable XL-RIS phase optimization. The DU framework unrolls a fixed number of
Riemannian gradient ascent steps on the sum-rate objective and recovers the discrete phase configuration via quantized phasor projection, learning only one step size per unrolled layer in a self-supervised manner without requiring AO-generated labels.

Numerical results showed that the DU framework achieves sum-rate performance close to that of the AO scheme while incurring substantially lower and nearly constant runtime across the considered system configurations. Both the AO and DU schemes consistently outperform the MO baseline under discrete hardware constraints.

The DU framework is trained offline under a fixed channel distribution and system configuration, and performance degradation may occur under significant distributional shifts. Future work will consider adaptive
extensions such as online fine-tuning, transfer learning, and meta-learning, as well as extensions to wideband near-field channels, mobility-aware environments, and distributed RIS control.

\bibliographystyle{IEEEtran}
\bibliography{references}

\vfill

\end{document}